\documentclass[openany]{nature}

\usepackage[T1]{fontenc}
\usepackage[left]{lineno}
\usepackage{amsthm,amsmath,amssymb}
\usepackage{mathrsfs}
\usepackage{overpic}
\usepackage{multirow}
\usepackage[T1]{fontenc}
\usepackage{graphicx}
\usepackage{pdflscape}
\usepackage{hyperref}
\usepackage{threeparttable}
\usepackage{threeparttablex}
\usepackage{xcolor}
\usepackage{ulem}
\usepackage{cancel}
\usepackage[figuresleft]{rotating}
\usepackage{caption}
\usepackage{subfigure}
\usepackage{tablefootnote}
\usepackage{hyperref}
\usepackage{graphicx}
\usepackage{longtable}
\usepackage{supertabular}
\usepackage{float} 
\usepackage{authblk}
\usepackage{multibib}
\usepackage{subfiles}
\usepackage{enumerate}
\usepackage{bm}
\usepackage{booktabs}

\hypersetup{
colorlinks=true,
linkcolor=red,
filecolor=blue,
urlcolor=cyan,
citecolor=green}

\makeatletter
 
\def\emph#1 {\textit{ #1 } }

\let\saved@includegraphics\includegraphics
\AtBeginDocument{\let\includegraphics\saved@includegraphics}
\renewenvironment*{figure}{\@float{figure}}{\end@float}
\makeatother

\newcommand{\araa}{Annu. Rev. Astron. Astrophys.}  
\newcommand{\aj}{Astron. J.}  
\newcommand{\apj}{Astrophys. J.}  
\newcommand{\apjl}{Astrophys. J. Lett.}  
\newcommand{\apjs}{Astrophys. J. Suppl. Ser.}  
\newcommand{\aap}{Astron. Astrophys.}  
\newcommand{\mnras}{Mon. Not. R. Astron. Soc.}  
\newcommand{\nat}{Nature} 
\newcommand{\prd}{Phys. Rev. D}  
\newcommand{\prl}{Phys. Rev. Lett.}  

\definecolor{dkblue}{RGB}{54, 86, 169}

\def\be{\begin{eqnarray}}
\def\ee{\end{eqnarray}}

\makeatletter

\title{Evidence for a brief appearance of gamma-ray periodicity after a compact star merger}

\author{Run-Chao Chen$^{1,2}$, Bin-Bin Zhang$^{1,2,3}$\thanks{E-mail: bbzhang@nju.edu.cn}, Chen-Wei Wang$^{4,5}$, Wen-Jun Tan$^{4,5}$, Shao-Lin Xiong$^{4}$\thanks{E-mail: xiongsl@ihep.ac.cn}, Jun Yang$^{1,2}$, Yi-Han Iris Yin$^{1,2}$, Shuang-Nan Zhang$^{4,5}$, Bing Zhang$^{6,7}$\thanks{E-mail: bzhang1@hku.hk}}

\begin{document}
\maketitle

\begin{affiliations}

\item School of Astronomy and Space Science, Nanjing University, 163 Xianlin Avenue, Nanjing 210023, People's Republic of China
\item Key Laboratory of Modern Astronomy and Astrophysics, Nanjing University, Ministry of Education, Nanjing 210023, People's Republic of China
\item Purple Mountain Observatory, Chinese Academy of Sciences, Nanjing 210023, People's Republic of China
\item State Key Laboratory of Particle Astrophysics, Institute of High Energy Physics, Chinese Academy of Sciences, Beijing 100049, People's Republic of China
\item University of Chinese Academy of Sciences, Chinese Academy of Sciences, Beijing 100049, China
\item Department of Physics, the University of Hong Kong, Pokfulam Road, Hong Kong, China
\item Nevada Center for Astrophysics and Department of Physics and Astronomy, University of Nevada Las Vegas, NV 89154, USA

\end{affiliations}

\clearpage

\begin{abstract}
The product of a compact star merger is usually hypothesized to be a hyperaccreting black hole, typically resulting in a gamma-ray burst (GRB) with a duration shorter than 2~s. However, recent observations of GRB~211211A and GRB~230307A, both arising from compact star mergers, challenge this model due to their minute-long durations. The data from both events are consistent with having a nascent, rapidly spinning highly magnetized neutron star (a millisecond magnetar) as the merger product and GRB engine, but a smoking gun signature is still missing. Here we report strong but not yet conclusive evidence for the detection of a 909-Hz gamma-ray periodic signal during a brief time window of GRB~230307A, which is consistent with the rotation frequency of such a millisecond magnetar. Notably, the periodic signal appeared for only 160~ms at an epoch coinciding with the transition epoch when the jet emission from the GRB central engine ceased and when the delayed emission from high latitudes started. If this signal is real, the temporal and spectral features of this gamma-ray periodicity can be consistently interpreted as asymmetric mini-jet emission from a dissipating Poynting-flux-dominated jet, as revealed by the energy-dependent light curve data of this burst.
\end{abstract}

Identifying whether the central engine of GRBs is a black hole or a neutron star frequently lacks clear direct evidence\cite{2014ARA&A..52...43B,Zhang_2018}. Although the gravitational wave event GW170817 confirmed that the pre-merger binary system was composed of two neutron stars\cite{2017PhRvL.119p1101A}, the characteristics of the resulting merger remnant are still undetermined. A distinguishing trait of a nascent neutron star, in contrast to a black hole, is its strong magnetic field and millisecond spin period. Identifying (quasi-)periodic signals in the emissions of GRBs and linking them to the central engine properties would be essential for discerning the nature of the central engine of GRBs\cite{1997ApJ...490..633K,2006A&A...454...11R,2010A&A...516A..16L}, particularly in the context of merger-type, mostly short-duration GRBs\cite{2005PhRvL..94t1101S,2014PhRvL.113i1104T,2021GReGr..53...59S,2013PhRvD..87h4053S}. However, the transient nature of GRB events, characterized by limited time series data and compounded by biases introduced by their non-stationarity\cite{Bachetti2022,2022ApJS..259...32H}, makes the detection of such periodic or quasi-periodic signals exceptionally challenging. Nonetheless, the recent identification of kilohertz quasi-periodic oscillations (QPOs) in two short GRBs\cite{2023Natur.613..253C}, demonstrated through broad power spectra features, has renewed optimism in this field. Motivated by these discoveries, our study aims to uncover further (quasi-)periodic signals in merger-type GRBs. In this endeavour, GRB~230307A has captured our focus. Associated with a kilonova\cite{2024Natur.626..737L,2024Natur.626..742Y}, this long-duration GRB strongly supports the compact star merger origin with a millisecond magnetar as its probable central engine\cite{2023arXiv230705689S}. As the second brightest GRB ever detected, the observed minute-long duration and substantial photon counts provide an ideal case for investigating (quasi-)periodicity.

To test the hypothesis that the central engine driving GRB~230307A is a nascent magnetar\cite{2023arXiv230705689S}, which may imprint periodicity at the spin frequency in its electromagnetic radiation, our objective is to identify this type of signal in the prompt emission phase. The observational data of GRB~230307A analysed in this study were obtained from the data archives of GECAM-B (Gravitational Wave High-energy Electromagnetic Counterpart All-sky Monitor satellites B)\cite{Li2021RDTM} and GECAM-C\cite{2023NIMPA105668586Z}, as well as the Fermi Gamma-ray Burst Monitor (Fermi/GBM)\cite{2009ApJ...702..791M}. A blind search for periodicity was conducted using event data from the three GECAM-B detectors with the smallest incident angles to the source. GECAM-B is prioritized because some of its detectors have incident angles of less than $40^{\circ}$. By contrast, GECAM-C and Fermi/GBM primarily have detectors with larger incident angles, typically greater than $50^{\circ}$. Furthermore, the Fermi/GBM data experienced saturation issues during certain time intervals\cite{2023GCN.33551....1D}, and the initial phase of the GECAM-C data was impacted by high-energy particle interference\cite{2023arXiv230705689S}, rendering these datasets unsuitable for a comprehensive blind search across the entire prompt emission phase.

To perform the blind search, we first applied a moving time window of 100~ms\cite{2022ApJS..259...32H, 2023Natur.613..253C}, which shifted by 7~ms at each step, corresponding to the minimum variability timescale, to the GECAM-B event data of GRB~230307A. This resulted in the extraction of 7,287 valid 100-ms segments (each with a signal-to-noise ratio, SNR$\gtrsim 10$) spanning from -0.021~s to 95.390~s relative to the trigger time $T_0$\cite{2023GCN.33406....1X} (see Methods for details). Each segment had count rates exceeding 4,000 photons per second and covered an energy range of 22--10,053~keV. To account for energy-dependent periodicity, each segment was further divided into subsets based on energy channels, resulting in 629,314 subsets of event data (see Methods). These subsets formed our final sample for the subsequent analysis.

\begin{figure}
\centering
\includegraphics[width=0.6\textwidth]{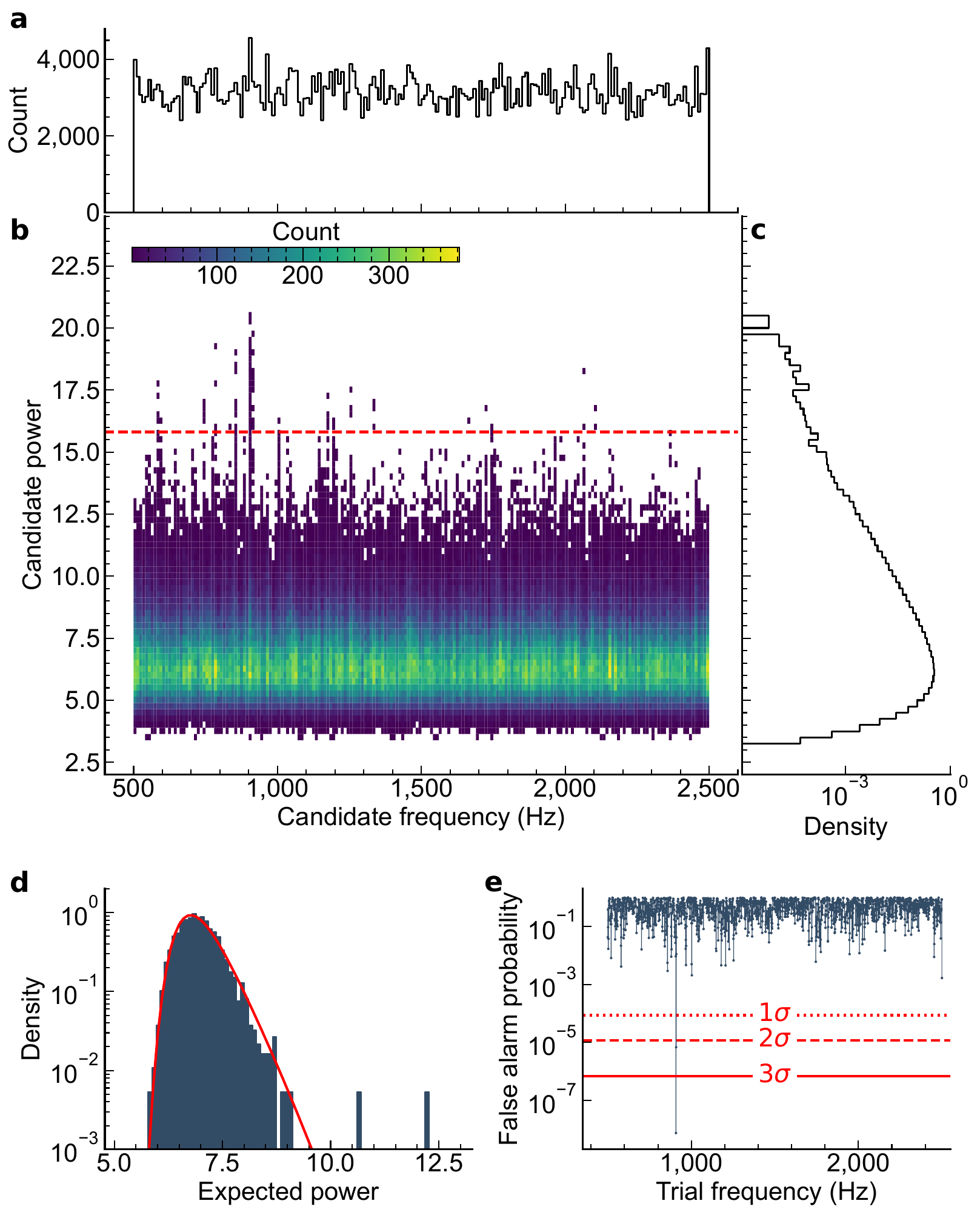}
\caption{\noindent\textbf{Results of blind search for periodicity in GRB~230307A.}
We obtained candidates from 629,314 subsets, covering the light curve within [-0.021, 95.390]~s from $T_0$ in the range 22--10,053~keV, using event data from three detectors with the smallest incident angles to the source onboard GECAM-B.
\textbf{a-c,} Joint and marginal distributions of candidate frequencies and powers. The joint distribution is displayed in \textbf{b}, while the marginal distributions of the candidate frequencies and powers are presented in panels \textbf{a} and \textbf{c}, respectively. The red dashed horizontal line indicates the threshold power $R_{\rm false} = 15.81$ calculated from the background searches (see Methods).
\textbf{d,} The distribution of the expected powers $R_{{\rm noise}, i}$ across different trial frequencies, assuming that all observed powers originate from the noise process (see Methods). The red curve shows the fitted probability density function for this distribution.
\textbf{e,} The single-trial FAP as a function of trial frequency, estimated from the asymptotic distribution shown in \textbf{d}. The red horizontal lines mark the $1\sigma$, $2\sigma$, and $3\sigma$ significance levels, accounting for the number of frequency trials conducted.}
\label{fig1}
\end{figure}

Using each subset of event data, we computed the Rayleigh power (see Equation~\ref{eq2}) across a trial frequency range of 500--2500~Hz with a resolution of 1~Hz. Each subset yielded a candidate, defined by the maximum Rayleigh power obtained from 2,001 trial frequencies. The trial frequency corresponding to this maximum power was designated as the candidate frequency, whereas the maximum power itself was labeled as the candidate power for that subset. Fig.~\ref{fig1}~\textbf{a--c} present the joint and marginal distributions of all candidate frequencies and powers obtained from all subsets. Under the null hypothesis that no periodicity exists, the ensemble of all candidate powers would follow a Gumbel distribution\cite{1928PCPS...24..180F,mood1950introduction}, comprising 629,314 variates. To account for the large number of searches, we derived a threshold power of $R_{\rm false} = 15.81$ from background data, which is used to screen out candidate powers that are unlikely to be observed based on a given number of searches. This corresponds to a tail probability $p = 1 - G(R_{\rm false}) = \frac{1}{629,314}$, where $G$ is the cumulative distribution function of candidate powers obtained from an equivalent blind search of background data. (see Methods and Equation~\ref{eq3}). We note that this $R_{\rm false}$, obtained from the analysis of background data, does not take into account rapid flux changes commonly observed in GRB light curves, and it should be taken as a lower limit to the threshold power for GRBs without QPOs. As shown in Fig.~\ref{fig1}~\textbf{b}, we identified 115 candidate powers exceeding $R_{\rm false}$ at 28 distinct candidate frequencies.

To determine the false alarm probability (FAP) associated with a candidate frequency $f_i$, we designed a three-step procedure: (1) Under the null hypothesis, the 629,314 candidate powers are expected to be uniformly distributed across 2,001 trial frequencies, leading to an expected average of $\frac{629,314}{2,001}$ candidate powers at any given frequency. Hence, we define a tail probability $p = \frac{1}{629,314 / 2,001}$ to represent the chance that a single candidate power at $f_i$ deviates significantly from the expected distribution purely due to noise. (2) For a given frequency $f_i$, if $N_{{\rm cand}, i}$ candidate powers are observed, we treat these as independent trials for the occurrence of false positives---cases in which noise alone produces a significant power. The typical noise-induced power level at $f_i$, denoted $R_{{\rm noise}, i}$, corresponds to a trial-corrected tail probability $\mathrm{Prob}_{{\rm noise}, i} = 1 - G(R_{{\rm noise}, i}) = (1 - p)^{N_{{\rm cand}, i}}$, where $G$ is the cumulative distribution function of all observed candidate powers at $f_i$ (see Methods). (3) By comparing the resulting $R_{{\rm noise}, i}$ values across all trial frequencies, we estimate the FAP associated with each candidate frequency, thereby assessing the statistical significance of each detected signal. 
Fig.~\ref{fig1}~\textbf{d} is the distribution of all $R_{{\rm noise}, i}$ values, calculated from the candidate power distributions across different $f_i$. Still under the null hypothesis, the ensemble of $R_{{\rm noise}, i}$ follows a generalized extreme value distribution\cite{1928PCPS...24..180F}. Thus, the single-trial FAP for any trial frequency $f_i$ can be derived from the distribution shown in Fig.~\ref{fig1}~\textbf{d}, which is further represented in Fig.~\ref{fig1}~\textbf{e}.

Because the blind search is performed across both the time and energy domains, these single-trial FAPs must be evaluated in the context of the total number of independent trials conducted in our analysis. However, this number is not straightforward to determine, as the 629,314 candidates were obtained from overlapping time segments and energy ranges, and the 2,001 trial frequencies were oversampled at 1~Hz intervals, much finer than the Rayleigh frequency resolution of the 100~ms segments ($\approx$10~Hz, given by $1/100~\text{ms}$). Based on our data segmentation strategy, we estimated the number of approximately independent trials as $(51~\text{s}/100~\text{ms}) \times (1/0.25) \times (2,000~\text{Hz}/10~\text{Hz}) \approx 4.08 \times 10^5$, where the three factors correspond to the number of non-overlapping time segments, non-overlapping energy ranges, and effective independent frequency bins, respectively.

After evaluating the FAPs of the 28 distinct candidate frequencies that yielded powers exceeding $R_{\rm false}$, we compared each single-trial FAP against the estimated value of $\frac{1}{4.08 \times 10^5} \approx 2.45 \times 10^{-6}$. We then corrected these probabilities for the number of variates in the distribution shown in Fig.~\ref{fig1}~\textbf{d}, which includes $R_{{\rm noise}, i}$ from 2,001 trial frequencies (Equation~\ref{eq9}). Following this procedure, we found that only the 908~Hz frequency demonstrated a statistically significant detection, with a trial-corrected $\mathrm{FAP}_{f_i = 908~\text{Hz}} = 1.45 \times 10^{-5}$. The 909~Hz frequency showed a marginal significance ($\mathrm{FAP}_{f_i = 909~\text{Hz}} = 1.34 \times 10^{-2}$), while all other frequencies were consistent with noise ($\mathrm{FAP} \approx 1$). Notably, all subsets yielding candidate powers exceeding $R_{\rm false}$ at 908 and 909~Hz overlapped within a 160~ms interval. This interval contained 68 subsets with candidate powers surpassing $R_{\rm false}$ in the 908--911~Hz range, indicating a potential QPO. The representative FAP of the QPO, defined as the minimum value of trial-corrected FAPs over the 908--911~Hz range, was approximately $1.45 \times 10^{-5}$. After integrating the overlaps, the QPO was identified (see Methods) at a peak frequency of 909~Hz, with the maximal $Z^2_1$ statistic\cite{1983A&A...128..245B,2021Natur.600..621C} derived from the time interval [24.401, 24.561]~s relative to $T_0$ in the 98--248~keV energy range. If we apply the estimated number of independent trials used in the blind search to correct for the detection of the signal's peak with $Z^2_1 \approx 52$ (corresponding to a single-trial chance probability of $5 \times 10^{-12}$ from $\chi^2_2$ distribution), the overall chance probability becomes $1 - (1 - 5 \times 10^{-12})^{(510\times4\times200)} \approx 2 \times 10^{-6}$. This FAP is consistent with the previous estimate, as the identified signal was detected within a 160~ms time interval, where the increased photon counts enhance the statistical significance.
We also note that in the subsequent application of the statistical tests, we report significance levels based solely on single-trial chance probabilities, without applying additional corrections for the number of trials.

\begin{figure}
\centering
\includegraphics[width=0.7\textwidth]{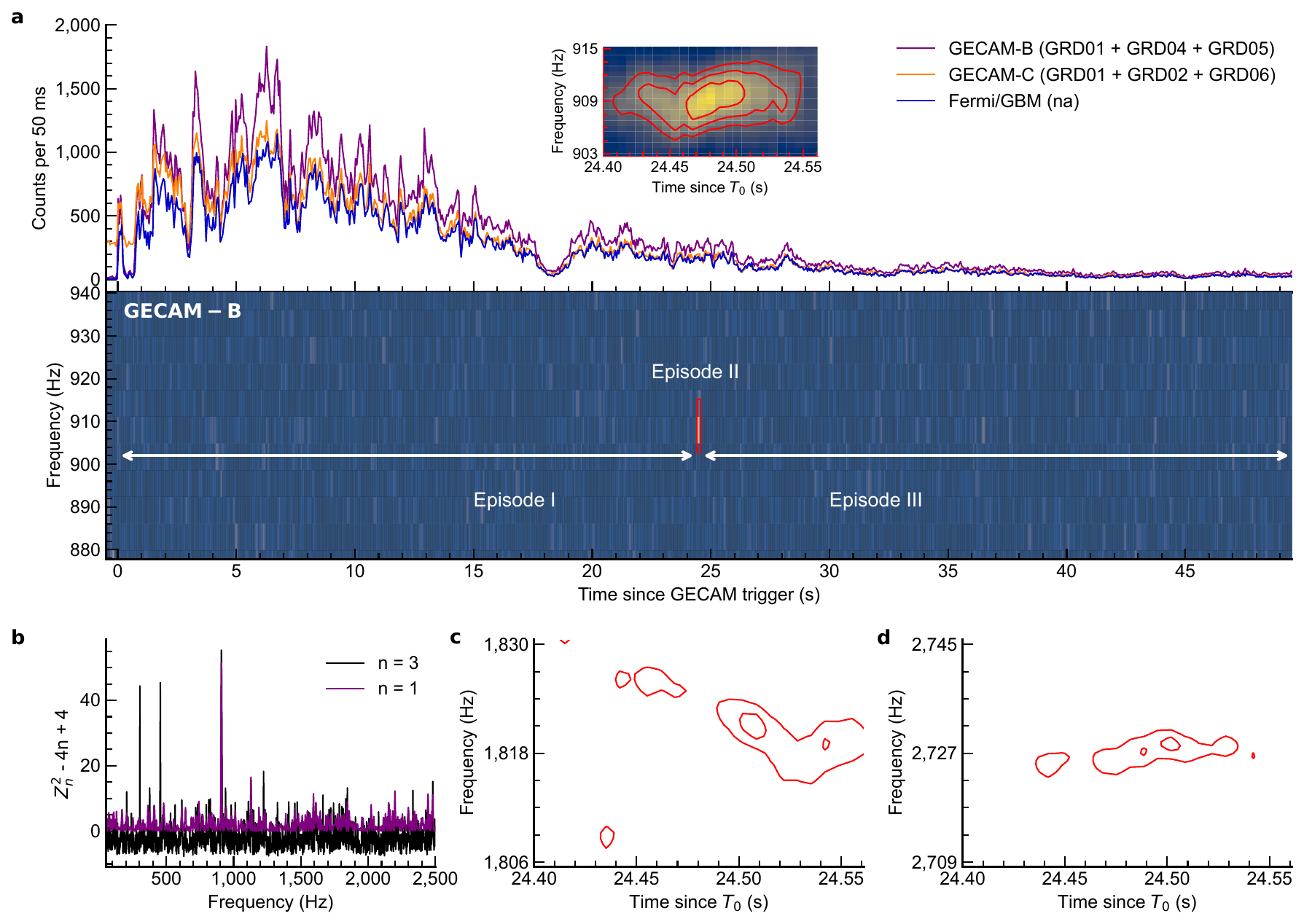}
\caption{\noindent\textbf{Occurrence and harmonics of the QPO signal in GRB~230307A.}
\textbf{a,} The location of QPO. The upper panel shows the light curves of GRB~230307A from various gamma-ray instruments in the energy range 98--248~keV. The lower panel shows the dynamical power spectra based on $Z^2_1$ statistics, calculated using overlapping 160-ms intervals with a step size of 7~ms. The frequency range analysed is $909\pm31.25$~Hz, with a resolution of 6.25~Hz.
A QPO signal, consistent with the periodicity identified in the blind search, is detected within the interval [24.401, 24.561]~s from $T_0$ at $909 \pm 6.25$~Hz. This detection is highlighted by the red box in the lower panel and further detailed in a zoomed-in view in the inset of the upper panel with an oversampling frequency resolution of 0.625~Hz. Red contours in the inset represent the $4\sigma$, $5\sigma$, and $6\sigma$ significance levels in $\chi^2_2$ distribution. Three distinct episodes are marked: ``Episode~I'', characterized by higher gamma-ray flux without a detectable QPO; ``Episode~II'', during which the QPO is observed; and ``Episode~III'', showing lower gamma-ray flux with no observable QPO.
\textbf{b,} Results of the $H$-test for the 909-Hz signal across various fundamental frequencies between 50 and 2,500~Hz, comparing $Z^2_1$ (purple) and $Z^2_3$ (black) statistics. Two subharmonics at 303~Hz and 405~Hz are evident in the $Z^2_3$ periodogram, indicating a strong presence of the 909-Hz signal. The higher peak at 909~Hz in $Z^2_3$ compared to $Z^2_1$ suggests the presence of two higher-order harmonics.
\textbf{c,} Dynamical $Z^2_1$ contours illustrating the second harmonic of the 909-Hz fundamental within the range $1818\pm12.5$~Hz. These contours, constructed from overlapping 160-ms intervals spaced by 7 ms, highlight the $95\%$ and $99\%$ significance levels in $\chi^2_2$ distribution.
\textbf{d,} Similar to \textbf{c}, but displaying the third harmonic of the 909-Hz fundamental within the range $2727\pm18.75$~Hz.}
\label{fig2}
\end{figure}

The location of this QPO signal is illustrated in Fig.~\ref{fig2}~\textbf{a}, where we conducted a cross-check using the dynamical power spectra\cite{2012ARA&A..50..609W} based on $Z^2_1$ statistics for event data from GECAM-B. According to the dynamical power spectra, the signal appears only between [24.401, 24.561]~s relative to $T_0$. Within this time interval, we performed the \(H\)-test\cite{1989A&A...221..180D} and identified two weak harmonics, observed at 1,818~Hz and 2,727~Hz (see Fig.~\ref{fig2}~\textbf{c-d}). Together with two other strong subharmonics observed at 303~Hz and 455~Hz, we confirmed that the fundamental of this signal is at 909~Hz, and estimated its root mean square (r.m.s.) fractional amplitude\cite{2005ApJ...634..547W} to be about $27.20^{+3.60}_{-3.64}\%$ within [24.401, 24.561]~s relative to \(T_0\) in the 98--248~keV range.
With knowledge of the specific time interval and energy range of the signal, we further searched (see Methods) for this 909-Hz signal from three different detectors onboard GECAM-B (Extended Data Fig.~\ref{efig1}). Our results suggest that such a signal can indeed be observed in these detectors. The single-trial chance probability estimated (see Methods) from the Monte Carlo simulation at 909~Hz is about $1.77 \times 10^{-8}$ when combining the data from these detectors (Extended Data Fig.~\ref{efig2}).

\begin{figure}
\centering
\includegraphics[width=\textwidth]{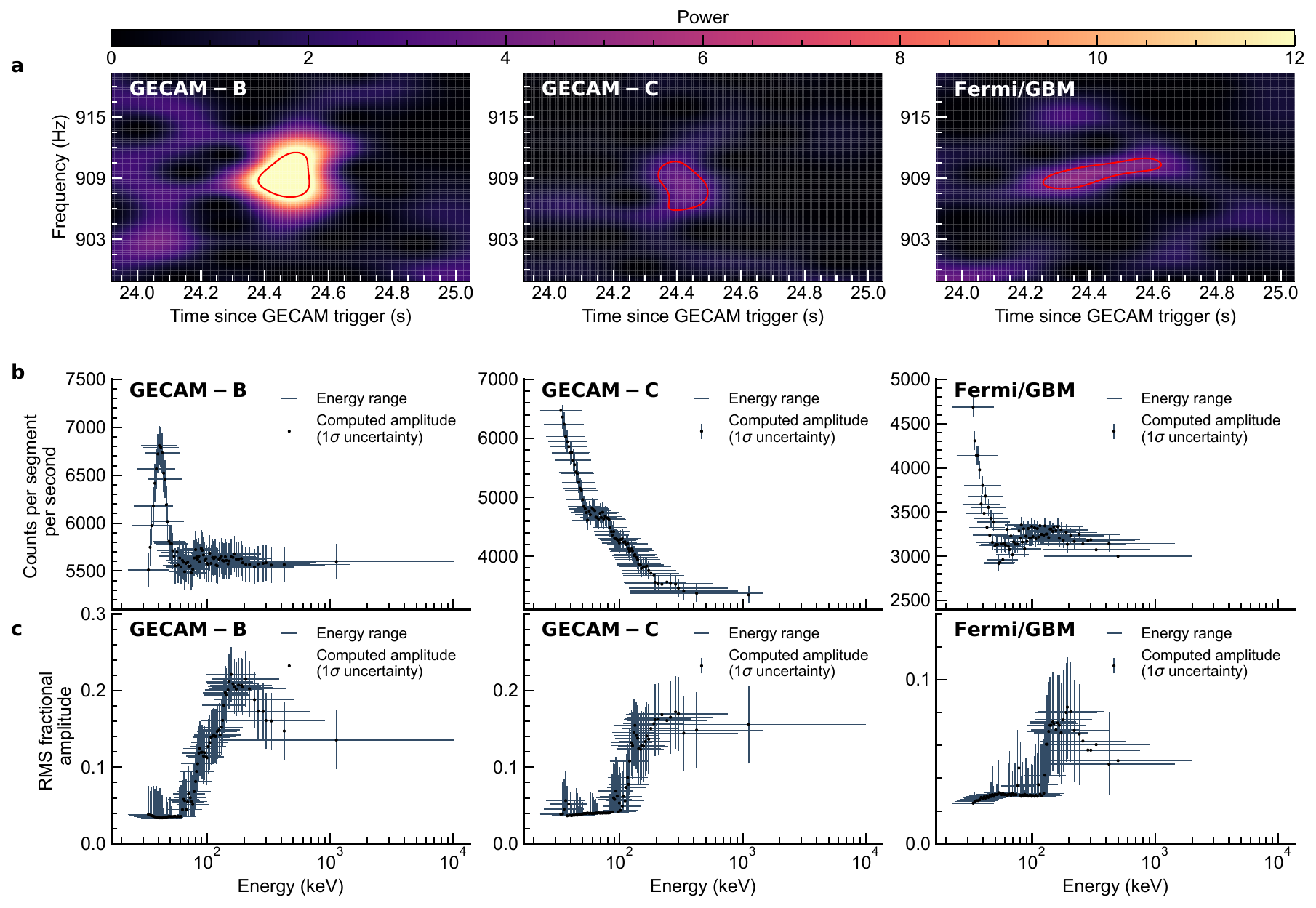}
\caption{\noindent\textbf{Multiple instruments' detections of consistent QPOs in GRB~230307A.} Data from GECAM-B and GECAM-C are obtained by combining three detectors with the smallest incident angles to GRB~230307A. Fermi/GBM data are extracted from the Sodium Iodide (NaI) detector na, as it is the only detector with an incident angle smaller than 60$^{\circ}$ to GRB~230307A.
\textbf{a,} WWZ spectrograms of light curves from different instruments, using a bin size of 0.1~ms. WWZ powers were calculated\cite{1996AJ....111..541F} with a time shift of 5~ms and a scale factor of \( 2\pi \times 100 \)~rad/s. The red contours represent the 97.76\% level of the local ensemble of WWZ powers from different instruments, corresponding to the 160-ms duration of the 909~Hz signal observed in the GECAM-B data.
\textbf{b,} Photon count rates across different energy ranges during the time intervals corresponding to the red contours in \textbf{a} for each instrument. Blue error bars represent measurements within these energy ranges (refer to Extended Data Table \ref{etab1} for details).
\textbf{c,} r.m.s. fractional amplitudes at 909~Hz. Blue error bars indicate measurements from the $Z^2_1$ statistics within the same time and energy intervals as in \textbf{b}. 
The QPO is observed simultaneously across detectors on different gamma-ray instruments, exhibiting consistent energy dependence. The strongest signal is detected by GECAM-B, which recorded a higher count rate than other instruments within the same energy range.}
\label{fig3}
\end{figure}

Furthermore, the presence of this QPO signal can be validated by examining its detection in gamma-ray data from other instruments. As the time interval of the signal does not overlap with the bad time interval of Fermi/GBM, and the GECAM-C background was less affected by high-energy particles during this period, we investigated the signal in observations from both GECAM-C and Fermi/GBM. Using the weighted wavelet Z-transform (WWZ) algorithm\cite{1996AJ....112.1709F}, we detected this signal with the GECAM-C and Fermi/GBM detectors within the same time window, although with much lower WWZ power (see Fig.~\ref{fig3}~\textbf{a}). This result confirms the validity of the signal observed by GECAM-B, which shows higher flux at smaller incident angles, leading to the highest r.m.s. fractional amplitude (see Fig.~\ref{fig3}~\textbf{b-c} and Extended Data Table~\ref{etab1}).

This QPO signal from GRB~230307A can also be confirmed by employing the methodology previously used to establish the existence of two other QPOs in GRBs~910711 and 931101B\cite{2023Natur.613..253C}. By fitting a Bayesian model to the power spectrum\cite{1975ApJS...29..285G} (see Methods), we computed a significant Bayes factor for the signal in GRB~230307A, similar to those found in the other two GRBs (Extended Data Fig.~\ref{fig3}). Notably, the signal detected in GRB~230307A, derived from the parameters that best fit a white noise model plus one Lorentzian, has a high coherence\cite{2000MNRAS.318..361N, 2012ARA&A..50..609W} of \( Q = 476 \) (see Methods). In contrast, the other signals in GRB~910711 (\( Q_1 = 22 \), \( Q_2 = 51 \)) and GRB~931101B (\( Q_1 = 29 \), \( Q_2 = 93 \)) show lower coherence\cite{2023Natur.613..253C}. Given that the power spectrum is computed from a 100-ms-length interval, which would naturally introduce a maximum coherence of \( Q = \frac{910}{10} = 91 \), the high coherence derived from the Bayesian model fitting indicates that the QPO observed in GRB~230307A is a genuine periodicity. The 1.1-ms spin period is fully consistent with that of a nascent millisecond magnetar, providing smoking-gun evidence for such a central engine in this source.

To further investigate the possible mechanism that could imprint such a rotational signal in the gamma-ray band, we analysed the spectral and temporal properties of this 909-Hz signal in GRB~230307A. First, by segmenting the data across different energy ranges while maintaining similar count rates over the same time interval of [24.401, 24.561]~s from $T_0$, we observed that the oscillation amplitude exhibits a consistent energy dependence across different instruments (GECAM-B, GECAM-C, and Fermi/GBM), showing an apparent diminishing trend at lower energy ranges (see Fig.~\ref{fig3}~\textbf{c}). Second, within the optimal energy range of 98--248~keV, we observed that the oscillation is observable only within a specific time interval, which aligns with the transition time\cite{2023arXiv230705689S} during which the prompt emission phase becomes dominated by the ``curvature effect''\cite{2000ApJ...541L..51K,2004ApJ...614..284D} related to delayed emission from high latitudes with respect to the line of sight (Extended Data Fig.~\ref{efig4}). These results imply that the 909-Hz signal exhibits distinct spectral and temporal characteristics, which help us further understand the emission geometry of the magnetar engine.

High-coherence oscillations in astrophysical transients related to neutron star spins have been widely observed in X-ray pulsars and thermonuclear X-ray bursts in accreting neutron star binary systems\cite{2000ARA&A..38..717V,2012ARA&A..50..609W}. However, unlike the X-ray bursts originating from the neutron star surface, the extreme brightness and isotropic energy of GRB~230307A\cite{2023arXiv230705689S} have conclusively constrained the origin of the 909-Hz signal to the relativistic jet launched from the magnetar engine. In order to have observed oscillations in jetted emission, three requirements must be met. First, the high spin of the central engine must be transported to the emission region. This implies the existence of a strong, twisted and ordered magnetic field in the system, with the magnetization parameter\cite{2011ApJ...726...90Z} $\sigma > 1$. Second, the emission region needs to be populated with ``hot spots'' with emissivity exceeding the surrounding region. A natural way of doing this is to introduce mini-jets with a local Lorentz factor $\gamma \approx \sqrt{\sigma + 1}$, which is approximately a few, in the comoving frame of the emitter. Third, the distribution of the hot spots needs to have axial asymmetry so that the spin of the engine can leave imprints on the emission region, causing a periodic modulation effect on the observed flux. This requires the number of randomly generated mini-jets not to be too large. We suggest that these conditions can be achieved within the framework of internal magnetic dissipation models involving a Poynting-flux-dominated flow, such as that envisaged in the internal-collision-induced magnetic reconnection and turbulence (ICMART) model of GRBs\cite{2011ApJ...726...90Z}. Such a model is already supported by the energy-dependent light curves of GRB~230307A\cite{2023arXiv231007205Y}, and the confirmation of a magnetar central engine for this burst provides the prerequisite for the highly magnetized outflow required for ICMART events to occur.

\begin{figure}
\centering
\includegraphics[width=\textwidth]{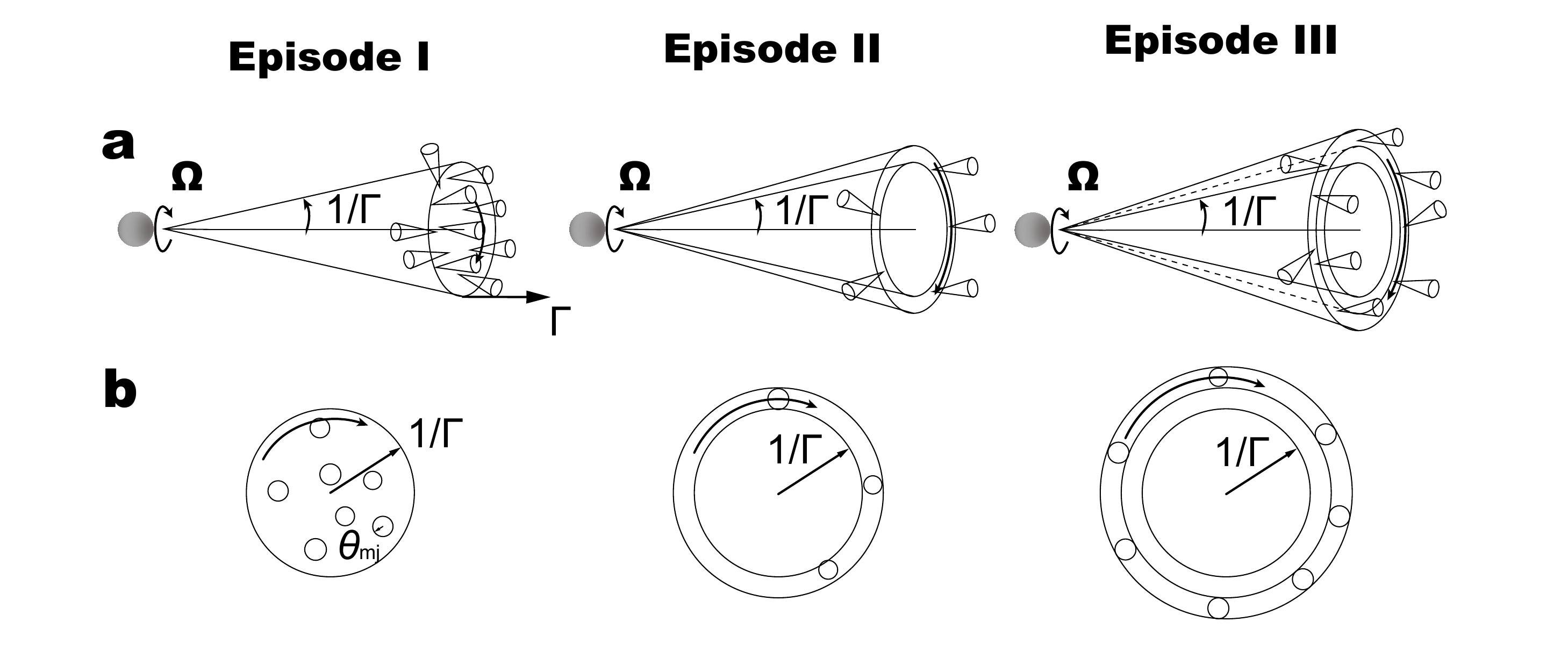}
\caption{\noindent\textbf{A physical scenario to interpret the 909-Hz signal in GRB~230307A.}
\textbf{a,} Schematic representation of GRB emissions during different episodes. The large cones illustrate the relativistic jet, which is causally connected with the magnetar central engine and is structured by an ordered helical magnetic field, forming a ``striped wind'' geometry\cite{2001A&A...369..694S}. 
Here, \(\Omega = 2\pi \times 909\)~rad/s represents the angular frequency of the magnetar central engine, which leaves imprints in the jet, and \(\Gamma\) denotes the bulk Lorentz factor of the jet in the observer's frame. Smaller cones within the jet depict mini-jets with half-opening angles of \(\theta_{\rm mj}\). The ensemble of mini-jets has a bulk Lorentz factor $\Gamma$, defining an emitting region with a half-opening angle of \(1/\Gamma\). Within this region, several bright mini-jets, oriented with half-opening angles $\theta_{\rm mj}$, have their emissions approximately beamed toward the observer. Mini-jets located at high latitudes, with half-opening angles \(\gtrsim 1/\Gamma\), reach the observer later, contributing to the HLE.
\textbf{b,} Projection of the emission region on the sky as seen by the observer. Episode~I: Quasi-symmetrically distributed mini-jets contribute to the observed gamma-ray emission without detectable periodicity. Episode~II: As the transition from the prompt emission to the HLE occurs, a small number of high-latitude mini-jets introduces asymmetry in their spatial distribution, enabling the detection of periodicity. Episode~III: At later times, the number of observed mini-jets increases, causing the HLE to become quasi-symmetrically distributed, rendering the periodicity undetectable again.}
\label{fig4}
\end{figure}

A physical scenario to interpret the brief 909-Hz signal from GRB~230307A is illustrated in Fig.~\ref{fig4}. The central engine of a millisecond magnetar launches a Poynting-flux-dominated wind, which dissipates at a radius $R$ with a bulk Lorentz factor of $\Gamma \gg 1$. In the emission region, many mini-jets with local Lorentz factor $\gamma$ are generated with different orientations in the local frame. Those pointing towards the observer (with a total Doppler factor of the order of ${\cal D} \approx \Gamma \gamma$) would make bright hot spots in the emission region. One can consider three episodes (Fig.~\ref{fig4}). Episode~I is the prompt emission phase, during which the observer can see a $1/\Gamma$ cone. Within this cone, there are likely many mini-jets with a half-opening angle of $\theta_{\rm mj} \approx 1/{\cal D} \approx (\Gamma\gamma)^{-1}$. These mini-jets are distributed with rough axial symmetry with respect to the line of sight. The rotation of the central engine would probably not leave an observable periodic modulation in the emission, so the periodicity was not detectable during this prompt emission phase. At a certain epoch, the prompt emission ceases abruptly. The observer starts to see emission from progressively higher latitudes with increasing ring sizes (see Fig.~\ref{fig4}). In the early high-latitude phase (Episode~II), the number of hot spots is relatively small (due to the small ring size), so that there is a likelihood of pronounced axial asymmetry. In this case, a periodic signal is observed right after the transition from prompt emission to the phase dominated by high-latitude emission (HLE). Later (Episode~III), the number of hot spots increases as the ring size increases. The emitter becomes roughly axially symmetric and periodicity can no longer be observed. This physical picture interprets the existence of the periodic signal during a brief time interval. Assuming that the duration of the 909-Hz signal corresponds to the time required for the mini-jet beam to sweep across the observer's line of sight due to its angular size, one can estimate the emission radius based on the duration $\Delta T \approx 0.16$~s, which gives $R \approx c \Delta T \theta_{\rm mj}^{-2} \approx (2\times 10^{15} \ \text{cm}) (\Delta T/0.16 \ \text{s}) (\Gamma\gamma/600)$, which is consistent with the expected radius from the ICMART model\cite{2011ApJ...726...90Z}. Since one can also write $R=\Gamma^2 c T$ (where $c$ is the speed of light, $T \approx 22$~s is the duration of the prompt emission phase\cite{2023arXiv230705689S}), one can further solve $\Gamma \approx 50$ and $\gamma \approx 12$ as the nominal parameters.

According to this model, the oscillation becomes more pronounced as the number of mini-jets drops. This is consistent with the observed spectral properties of the 909-Hz signal (see Fig.~\ref{fig3}~\textbf{c}). The observed diminishing trend of the oscillation at lower energies can be explained because low-energy photons originate from a larger number of mini-jets with a wider range of orientations, which smooths out the asymmetry. By contrast, higher-energy photons can only be observed from the mini-jets oriented towards the observer, preserving the asymmetry and enhancing the oscillation amplitude. Additionally, the decay of the r.m.s. amplitude with the inclusion of even higher-energy photons is expected, as the synchrotron radiation spectra of mini-jets naturally lead to a decrease in photon numbers beyond the cutoff energy, typically around the mega-electronvolt range\cite{2023arXiv230705689S}.

We also performed a blind search for GRB~211211A using Fermi/GBM data (see Methods). The most possible QPO was identified at 935~Hz, which is close to the 909-Hz spin frequency detected in GRB~230307A. However, its FAP of approximately $1.09\%$ renders this detection statistically insignificant (Extended Data Fig.~\ref{efig5}). GRB~211211A is also a very bright long-duration GRB associated with a kilonova\cite{2022Natur.612..223R,2022Natur.612..228T,2022Natur.612..232Y}, and two low-frequency QPOs have been reported for this burst by refs.\cite{2024ApJ...970....6X,2024ApJ...967...26C}. However, unlike GRB~230307A that shows a well-defined energy-dependent single broad component with a clear signature of HLE\cite{2023arXiv230705689S}, it has several emission episodes and does not show evidence of HLE\cite{2022Natur.612..232Y}. Physically, its multi-episode duration is probably defined by the progenitor or engine\cite{2025JHEAp..45..325Z}. Based on these observations, we conclude that the conditions required to generate an observable kilohertz periodicity in GRBs are stringent and the cases should be rare. These conditions include the brightness of the burst, whether the light curve includes episodes of HLE, and, if it does, whether the hot spots have an asymmetric distribution during the early phase of HLE. In any case, our model predicts that in rare cases, where a small number of observable mini-jets persist throughout the entire prompt emission phase, a (quasi-)periodic signal may be detected. This could explain the previously reported QPOs in GRB~910711 and GRB~931101B\cite{2023Natur.613..253C}, where fewer mini-jets during the prompt emission phase may have led to a detectable (quasi-)periodicity, albeit with lower power at the peak frequency.

\captionsetup[table]{name={\bf Table}}
\captionsetup[figure]{name={\bf Fig.}}

\clearpage

\clearpage

\clearpage

\clearpage

\clearpage
\section*{Methods}
\subsection{Search for Periodicity in GRB~230307A\\}
\noindent\textbf{Data Acquisition and Processing.} The prompt emission data for GRB~230307A were accessed from the data archive of GECAM-B, GECAM-C, and Fermi/GBM. Given that the data quality of Fermi/GBM and GECAM-C, due to the presence of bad time intervals and incident angles to the source, is not suitable for a full blind search, we focused solely on the GECAM-B data for the search for periodicity while checking for consistency with the data from GECAM-C and Fermi/GBM if any detected signal occurs during the good time intervals of these instruments. For the selection of detectors, instead of simply selecting the most illuminated detectors\cite{2023Natur.613..253C}, or attempting different combinations of all detectors until the most significant signal is identified\cite{2025MNRAS.tmp..154Y}, in this study, considering that the search for periodicity in GRB~230307A accounts for energy dependence, and that detectors with large incident angles may record events where photons have undergone substantial scattering within the spacecraft body before reaching the detectors\cite{2011arXiv1111.0514W}, we restrict our analysis to data from at most three detectors with the smallest incident angles to the source and ensure that each selected detector has an incident angle of less than $60^{\circ}$. Specifically, for GECAM-B, we use GRD01, GRD04, and GRD05; for GECAM-C, we use GRD01, GRD02, and GRD06; and for Fermi/GBM, we use only the NaI detector na.

To determine the duration of the prompt emission and the minimum variability timescale (MVT)\cite{2015ApJ...811...93G,2023A&A...671A.112C} of GRB~230307A, we applied the Bayesian block algorithm\cite{2013ApJ...764..167S} to the combined event data from the GECAM-B detectors GRD01, GRD04, and GRD05, which cover the energy range of 22--10,053~keV. The Bayesian block duration was calculated to be [-0.021, 95.390]~s from $T_0$, and the MVT was calculated to be 7 ms.

The event data were first sliced into segments using a moving time window of 100~ms\cite{2022ApJS..259...32H, 2023Natur.613..253C}, with a step size of $\Delta t^\prime$, applied to the GECAM-B event data of GRB~230307A. The 100-ms segmentation duration was chosen as it was sufficient to capture quasi-periodic signals if the remnant powering GRB~230307A were a hypermassive neutron star, similar to the QPOs detected in GRB~910711 and GRB~931101B\cite{2023Natur.613..253C}.

To determine the shifting step $\Delta t^\prime$, we first require $\Delta t^\prime < 100$~ms to ensure overlapping time windows. The use of overlapping time windows minimizes the risk of missing quasi-periodic or periodic signals that briefly appear near the edge of a window and might not be fully captured within a single 100-ms segment. Second, we consider the following scenario. If a periodic signal persists for a duration longer than $(100 + \Delta t^\prime)$~ms, the signal will be captured multiple times by the overlapping 100-ms time windows advancing with a step of $\Delta t^\prime$. This repeated capture leads to multiple detections of extreme power at the same frequency $f$. Consequently, the distribution of extreme power values at $f$ will deviate from the theoretical distribution expected under the null hypothesis (i.e., assuming no periodicity exists throughout the entire duration of the GRB). This deviation enables the identification of a potential periodic signal. However, within a specific time interval $[t_1, t_1 + 100]$~ms, if an extreme power is detected at frequency $f$ but arises solely due to the large number of searches rather than a periodic signal, we conservatively assume that such noise-driven signals should not persist in the subsequent interval $[t_1 + \mathrm{MVT}, t_1 + \mathrm{MVT} + 100]$~ms. This assumption is based on the fact that the MVT serves as an indicator of the shortest timescale over which statistically significant variations exceed the noise level in the GRB’s temporal profile. Therefore, we adopt the MVT as the step size $\Delta t^\prime$. Even if a genuine periodic signal generates extreme power in one 100-ms interval, if it disappears in the subsequent interval, it would be indistinguishable from a noise-induced false signal that also produces extreme power due to the large number of searches.

Further segmentation based on energy is designed to account for the dependence of potential periodicity on photon energy. To ensure that the selected segments contain a substantial fraction of source photons rather than background noise and are bright enough to support further energy segmentation, we first fit a third-order polynomial to the light curve using a bin size of 100~ms over the background time intervals of $[-100, -0.021]$~s and $[95.390, 200]$~s relative to $T_0$. We then calculate the signal-to-noise ratio (SNR) for each segment, assuming Poisson-distributed data with a Gaussian background\cite{2018ApJS..236...17V}. Only segments with $\text{SNR} \geq 10$ are selected for further analysis, yielding subsets totaling about 51~s in duration.

Furthermore, given the rapid spectral evolution of the GRB and the flux variability across different segments (see, e.g., ref.\cite{2023arXiv230705689S}), we required a consistent fraction of events contained in each subset derived from its parent segment. Specifically, each subset of event data is defined by:
\begin{equation}
\mathcal{S} \in \lbrace t_{\rm ph}, E_{\rm ph} \mid t_{\rm ph} \in \mathcal{T}_{\text{segment}}, E_{\rm ph} \in \mathcal{E}_{\text{subset}}, \mathcal{F}_{\text{ph}} \gtrsim 0.25 \rbrace,
\label{eq1}
\end{equation}
where $t_{\rm ph}$ and $E_{\rm ph}$ represent the arrival time and energy channel of each event, respectively. $\mathcal{T}_{\text{segment}}$ represents the time interval of the parent segment, $\mathcal{E}_{\text{subset}}$ represents the consecutive energy channel range of each subset, $\mathcal{F}_{\text{ph}}$ represents the fraction of events in each subset relative to those in its parent segment.

For example, we consider a segment within the time interval $\mathcal{T}_1$, where the lowest-energy photon corresponds to energy channel $ch_1$ and the highest-energy photon to channel $ch_n$. To determine the first subset, we set $ch_1$ as the lower boundary of the energy channel range and progressively increase the upper boundary, calculating $\mathcal{F}_{\text{ph}}$ at each step, starting from $ch_1 + 1, ch_1 + 2$, and so on. This process continues until we find that $\mathcal{F}_{\text{ph}} < 0.25$ for the energy range $[ch_1, ch_1 + x]$ but $\mathcal{F}_{\text{ph}} \geq 0.25$ for $[ch_1, ch_1 + x + 1]$. At this point, the first subset is determined as 
$\mathcal{S}_1 = \{t_{\text{ph}}, E_{\text{ph}} \mid t_{\text{ph}} \in \mathcal{T}_1, E_{\text{ph}} \in \mathcal{E}_1 = [ch_1, ch_1 + x + 1]\}$. Subsequent subsets are formed by shifting the lower boundary to $ch_1 + 1$, $ch_1 + 2$, and so forth, repeating the process until the upper boundary of the final subset reaches $ch_n$. The threshold fraction of 0.25 was chosen based on the observation that a 100-ms segment with $\text{SNR} = 10$ typically contains approximately 400 counts. To ensure that each subset has at least 100 counts, providing sufficient statistics for a reliable periodicity test, we set the energy segmentation fraction to $\frac{100}{400} = 0.25$.

\noindent\textbf{Methodology.}
After segmenting the event data based on time and energy, we obtained $N_{\rm subset}$ subsets from $N_{\rm segment}$ segments. To perform a blind search for periodicity across these subsets, we first identified an ensemble of candidates by applying the Rayleigh test\cite{10.1111/j.2517-6161.1975.tb01550.x} to each subset. We selected the Rayleigh test because it operates directly on the event data without requiring binning into a uniform time series, thereby avoiding the subjectivity and potential sensitivity loss associated with binning methods\cite{2002ApJ...576..932K}. For instance, if the frequency range extends up to the Nyquist frequency $f_{\max}$, the required bin size would be $\frac{1}{2f_{\max}}$. This necessitates resampling the recorded event data into discrete bins, leading to a biased selection of centroid times, with an associated error of $\pm \frac{1}{2f_{\max}}$. To mitigate these biases, we compute power directly from the event data rather than relying on binned representations.

For each subset of event data, a candidate frequency $R_x$, where $x = 0, 1, \ldots, N_{\rm subset} -1$ was selected based on the maximal Rayleigh power (i.e., the candidate power) across a series of trial frequencies, and the trial frequency corresponds to $R_x$ was labeled as $f_x$ (i.e., the candidate frequency). The Rayleigh power at frequency \( f \) is calculated as follows:
\begin{equation}
R(f) = \frac{1}{N}\left[ \left(\sum^N_{j=1}\cos(2\pi f t_j)\right)^2 + \left(\sum^N_{j=1}\sin(2\pi f t_j)\right)^2 \right],
\label{eq2}
\end{equation}
where \( N \) is the total number of photons arriving at times \( t_j \) relative to $T_0$. This formula is also known as the \( Z^2_1 \) statistic\cite{1983A&A...128..245B} but differs by a factor of 2.

The frequency range spanned from 500 to 2,500~Hz. The lower limit was chosen to filter out low-frequency power excess due to the red noise process\cite{2023Natur.613..253C}, and the upper limit was selected to encompass both the Keplerian frequency and the rotational breakup frequency of a remnant neutron star with an assumed mass of approximately 2.7~$M_{\odot}$ and a radius of around 12~km from the merger. 
An oversampling factor of 10 was applied, meaning that we uniformly sampled 10 points around each Fourier frequency. Finally, all trial frequencies could be marked as $f_i$, where $i = 0, 1, \ldots, 2000$ and $f_i = 500, 501, \ldots, 2500$~Hz, and the trial number of frequency in this study shall be $N_{\rm freq} = 2{,}001$.

Assuming no periodicity exists in all extracted subsets (the null hypothesis), the observed candidate power $R_x$ at a candidate frequency $f_x$ from a subset would represent an extreme value, specifically the maximum drawn from $N_{\rm freq}$ random variates of the exponential distribution with an expectation of 1. According to the Fisher-Tippett theorem\cite{1928PCPS...24..180F}, the ensemble of $R_x$ follows the Gumbel distribution\cite{mood1950introduction}:
\begin{equation}
G(R_x) = e^{-e^{-(\frac{R_x-\mu}{\beta})}},
\label{eq3}
\end{equation}
where $G(R_x)$ is the cumulative distribution function of $R_x$, and \( \mu \) and \( \beta \) are the location and scale parameters, respectively. Using the method of moments, we compute \( \mu \) and \( \beta \) as follows:
\begin{equation}
{\rm E}[R_x] = \mu + \gamma\beta; \quad {\rm V}[R_x] = \frac{\pi^2}{6} \beta^2,
\label{eq4}
\end{equation}
where ${\rm E}[\cdot]$ and ${\rm V}[\cdot]$ denote the expectation and variance, respectively, and \( \gamma\simeq 0.577216 \) is the Euler's constant.

\textbf{1. Using the background search to determine the threshold power.} Under the null hypothesis, the ensemble of candidate powers from all subsets follows a distribution with $N_{\rm subset}$ variates drawn from the Gumbel distribution. Given these $N_{\rm subset}$ trials, the threshold power corresponding to a tail probability of $p = \frac{1}{N_{\rm subset}}$ is expected, on average, to be exceeded by one candidate power purely by chance. Thus, the threshold power can be determined from this probability using a simulated distribution of candidate powers obtained from data without periodicity. To achieve this, we conducted a blind search for periodicity using background data to generate the simulated distribution.

If the burst duration is defined as $[t_1, t_1 + \Delta T_{\rm burst}]$, the background data duration is set as $[t_1 - \Delta T_{\rm burst}, t_1]$. We applied the same segmentation strategy described earlier to obtain subsets of event data from the background, using the same detectors as those used for burst data extraction. However, due to the lower photon count rates in the background segments, further energy-based segmentation was not feasible. Therefore, each background segment was treated as a single subset, resulting in $N_{\rm segment}$ background subsets, equal to the number of segments in the burst duration.

To obtain the simulated distribution, we calculated candidate powers from the background subsets. Since the background data should not contain any periodicity, the ensemble of all candidate powers from the background subsets forms a distribution with $N_{\rm segment}$ variates drawn from the Gumbel distribution. However, because the sample sizes of the candidate power distributions from the burst and the background differ, we corrected the tail probability of the threshold power for the simulated distribution as
\begin{equation}
\mathrm{Prob}_{\rm false} = 1 - (1 - p)^{r} \approx \frac{1}{N_{\rm segment}},
\label{eq5}
\end{equation}
where $r = \frac{N_{\rm subset}}{N_{\rm segment}}$. This probability adjustment ensures that $r$ iterations of background searches are treated as equivalent to the blind search conducted on the burst data. That is, given a single-trial chance probability $p$, we expect to observe one false positive over $r$ repetitions of background-based searches, each conducted across $N_{\rm segment}$ subsets. Consequently, using the parameters of the simulated distribution obtained from the background search, we calculate the threshold power as
\begin{equation}
R_{\rm false} = \mu - \beta \ln[-\ln (1 - \mathrm{Prob}_{\rm false})].
\label{eq6}
\end{equation}

\textbf{2. Calculating the FAP for potential periodicity.} Once a candidate from a subset exhibits $R_x > R_{\rm false}$, the corresponding candidate frequency $f_x$ is considered a potential periodicity frequency. Since there are only $N_{\rm freq}$ trial frequencies, all candidate frequencies $f_x$ belong to one of these $N_{\rm freq}$ discrete frequency bins. The FAP that the potential periodicity occurs at $f_x$ shall be the probability that the trial frequency $f_i = f_x$ stands out from all $N_{\rm freq}$ trial frequencies.

Under the null hypothesis, all trial frequencies $f_i$ should have an equal probability of producing a false positive detection. Thus, for $N_{\rm subset}$ candidates uniformly distributed across $N_{\rm freq}$ trial frequencies, the tail probability that a candidate power at any specific trial frequency $f_i$ deviates significantly from the noise-induced distribution is given by $p = \frac{1}{N_{\rm subset}/N_{\rm freq}} = \frac{N_{\rm freq}}{N_{\rm subset}}$. As more subsets are evaluated, the total number of candidates increases, leading to a lower tail probability at each frequency. This, in turn, implies a higher expected maximal power due to chance alone, thereby elevating the risk of a false positive detection.

For a given frequency $f_i$, the number of associated candidates, $N_{{\rm cand}, i}$, represents the number of independent trials at that frequency. That is, we have effectively conducted $N_{{\rm cand}, i}$ independent searches at $f_i$. Consequently, the probability that none of the $N_{{\rm cand}, i}$ trials at $f_i$ yields a candidate power deviating from the noise-induced distribution is
\begin{equation}
\mathrm{Prob}_{{\rm noise}, i} = (1 - p)^{N_{{\rm cand}, i}},
\label{eq7}
\end{equation}
which we adopt as the tail probability for estimating the expected maximum power arising from noise at $f_i$. As this probability decreases with increasing $N_{{\rm cand}, i}$, the expected maximal power from the noise distribution rises accordingly, implying that any observed excess becomes less likely to be explained by noise alone.

Using the parameters of the distribution of $N_{{\rm cand}, i}$ candidate powers, we can estimate the expected power due to noise at $f_i$ as
\begin{equation}
R_{{\rm noise}, i} = \mu_{f_i} - \beta_{f_i} \ln\left[-\ln \left(1 - \mathrm{Prob}_{{\rm noise}, i}\right)\right].
\label{eq8}
\end{equation}
According to the Fisher-Tippett theorem, the ensemble of $R_{{\rm noise}, i}$ values across all trial frequencies follows a generalized extreme value distribution. We use the \texttt{scipy.stats} package in \texttt{Python} to approximate this distribution and compute the single-trial FAP for each frequency $f_i$.

To determine the probability that a specific trial frequency $f_i = f_x$ stands out among all $N_{\rm freq}$ trial frequencies, we apply a correction to the single-trial probability as follows:
\begin{equation}
{\rm FAP}_{f_i = f_x} = 1 - \left[1 - p(f_i)\right]^{N_{\rm freq}},
\label{eq9}
\end{equation}
where $p(f_i)$ represents the single-trial FAP for $f_i$.

\noindent\textbf{Results.}
In this study, we searched for periodicity in GRB~230307A using $N_{\rm subset} = 629,314$ subsets derived from $N_{\rm segment} = 7,287$ time segments. The joint distribution of all candidate frequencies and powers is presented in Fig.~\ref{fig1} \textbf{a-c}. The marginal distributions of candidate frequencies and powers exhibit no apparent outliers, indicating that the periodic signal is not consistently present throughout the burst. This outcome is expected and supports the validity of our segmentation strategy.

In our background search, we obtained a threshold power of \( R_{\rm false} = 15.81 \) and identified 115 subsets with candidate powers exceeding $R_{\rm false}$ at 28 distinct candidate frequencies. After correcting for the number of trial frequencies, $N_{\rm freq} = 2{,}001$, only the candidate frequencies 908~Hz and 909~Hz yielded ${\rm FAP}_{f_i = \text{908~Hz}} = 1.45 \times 10^{-5}$ and ${\rm FAP}_{f_i = \text{909~Hz}} = 1.34 \times 10^{-2}$, whereas all other candidate frequencies resulted in ${\rm FAP} \sim 1$ (Fig.~\ref{fig1} \textbf{e}).

Additionally, we found that all 38 subsets yielding candidate powers exceeding $R_{\rm false}$ at 908~Hz overlap with all 18 subsets yielding candidate powers exceeding $R_{\rm false}$ at 909~Hz within a 160~ms time interval. This suggests that the periodicity detected at both frequencies may originate from the same QPO signal. Furthermore, this 160~ms interval also coincides with all 6 subsets yielding candidate powers exceeding $R_{\rm false}$ at 910~Hz, as well as all 6 subsets yielding candidate powers exceeding $R_{\rm false}$ at 911~Hz. Since the FAPs at 910 and 911~Hz (${\rm FAP}_{f_i = \text{910~Hz}}$ and ${\rm FAP}_{f_i = \text{911~Hz}}$) are close to 1, we conclude that this QPO likely peaks around 908 or 909~Hz and exhibits very high coherence.

Since the signals at 908 and 909~Hz originate from subsets within the same time interval, and a 100~ms segment naturally introduces a Rayleigh frequency resolution of 10~Hz, the frequency range 908--911~Hz is therefore both physically and statistically relevant. To further determine a representative FAP of this QPO signal, we adopt the minimum trial-corrected FAP among the 908--911~Hz range as a conservative estimate of the QPO's statistical significance:
\begin{equation}
{\rm FAP} = \min_j \lbrace { \rm FAP}_{f_i = f_j} \rbrace,
\label{eq10}
\end{equation}
where $f_j$ represents the trial frequencies within the 908--911~Hz range. Finally, we obtained a representative FAP of approximately $1.45\times 10^{-5}$ (equivalent to $4.2\sigma$) for this QPO, indicating that it is a statistically significant signal that is unlikely to arise from random noise.

\subsection{Validation of the QPO in GRB~230307A\\}

\noindent\textbf{Optimal Interval of the QPO Based on the $Z^2_1$ Search.} 
After combining the time intervals of all subsets with candidate power exceeding the $R_{\rm false}$ threshold within 908--911~Hz, we identified a 160-ms interval occurring between [24.404, 24.564]~s relative to $T_0$. To determine the optimal energy range for the QPO, we calculated the $Z_1^2$ statistic across different combinations of energy channels from the time interval within [24.404, 24.564]~s from $T_0$. Specifically, we first considered the full energy range (22--10,053~keV, corresponding to channels [37, 447]) and calculated the $Z_1^2$ statistic at the frequency $909 \pm 1/0.16$~Hz with a resolution of $0.1/0.16$~Hz, obtaining the maximum $Z_1^2$ statistic $P_0$. We then iteratively reduced the high-energy limit, calculating the maximum $Z_1^2$ statistic $P_n$ at different channel ranges [37, $447 - n$] for $n = 1, 2, \dots$, and identified the maximum $Z_1^2$ statistic $P_x = \max\{P_n\}$. Subsequently, we adjusted the low-energy limit, calculating the maximum $Z_1^2$ statistic $P_m$ at different channel ranges [$37 + m$, $447 - x$] for $m = 1, 2, \dots$, to find the maximum $Z_1^2$ statistic $P_y = \max\{P_m\}$. The final optimal energy range was determined as [$37 + y$, $447 - x$] = [99, 168], corresponding to approximately 99--249~keV, where the signal exhibited the highest $Z_1^2$ statistic.

It should be noted that, while no other consecutive energy channels yield a $Z^2_1$ statistic larger than the energy channel range 99--168, this may not necessarily represent the most restrictive range in which the QPO appears. The key observation is that the signal-to-noise ratio of the QPO peaked at 909~Hz decreases when the energy range is either expanded or narrowed relative to this range. For instance, in the ranges corresponding to channels 98--169, 99--169, 100--168, or 99--167, the maximal $Z^2_1$ statistic still peaks at 909~Hz but decreases compared to its value in the 99--168 range.

Furthermore, considering a potential systematic uncertainty of 7~ms in the start time of the segmented light curve, we employed a segment length of 160~ms with a step length of 1~ms. By adjusting the start time of the QPO within \(24.404 \pm 0.007\)~s from $T_0$, we identified the optimal 160-ms interval. The maximal \(Z^2_1\) statistic at 909~Hz was observed within the duration of [24.401, 24.561]~s from $T_0$.

\noindent\textbf{Harmonic Analysis.}
The \(Z^2_n\) statistic, a generalized form of the \(Z^2_1\) statistic, is used for identifying periodic signals that may include harmonics. It is defined as:
\begin{equation}
Z^2_n = 2\sum^n_{k=1} R(kf),
\label{eq11}
\end{equation}
where \(R(kf)\) is the Rayleigh power at the $k$-th harmonic of the frequency \(f\). To determine the optimal number of harmonics for a given frequency \(f\), we employed the \(H\)-test\cite{1989A&A...221..180D}, which is calculated as:
\begin{equation}
H = \max(Z^2_m - 4m + 4),
\label{eq12}
\end{equation}
where \(m\) ranges from 1 to 20. For the 909-Hz signal in GRB~230307A, we found the optimal harmonic number to be 3, with an \(H\) statistic of approximately 55.32. This is illustrated in Fig. \ref{fig2} \textbf{b-d}, where two weak signals above the 99\% significance level in \(\chi^2_2\) were observed at 1,818 Hz and 2,727 Hz, consistent with the fundamental frequency of 909 Hz.

\noindent\textbf{Computation of the Fractional Amplitude.}
Given the identified 909-Hz fundamental frequency, along with the optimal time interval and energy range, the r.m.s. fractional amplitude of the QPO can be computed using the \(Z^2_n\) statistic, as described in ref.\cite{2005ApJ...634..547W}:
\begin{equation}
r = \left(\frac{\hat{Z^2_n}}{N_{\gamma}}\right)^{\frac{1}{2}}\left(\frac{N_{\gamma}}{N_{\gamma} - N_b}\right),
\label{eq13}
\end{equation}
where \(N_{\gamma} = 893\) represents the total number of photons within the optimal interval for the 909-Hz signal. \(N_b = 53\) is the number of background photons, estimated by fitting a polynomial to the light curve of GRB~230307A in the background regions spanning [-100, -10]~s and [100, 300]~s from \(T_0\) within the 98--248~keV range. The term \(\hat{Z^2_n}\) denotes the real signal power, with \(n = 3\) as determined by the \(H\)-test.

To estimate the real signal power, we assume the null hypothesis that the measured power \(Z^2_3\) originates from a signal with a given power \(\hat{Z^2_3}\). The probability of observing a power \(Z^2_n\) given the expected power \(\hat{Z^2_n}\) is described by\cite{1975ApJS...29..285G,1994ApJ...435..362V}:
\begin{equation}
p_n(Z^2_n \mid \hat{Z^2_n}) = \left(\frac{Z^2_n}{\hat{Z^2_n}}\right)^{(n-1)/2} e^{-(Z^2_n + \hat{Z^2_n})/2} I_{n-1}(\sqrt{Z^2_n \hat{Z^2_n}}),
\label{eq14}
\end{equation}
where \(I_{n-1}\) is the modified Bessel function of the first kind. Utilizing the \texttt{Python} package \texttt{scipy.stats}, we modeled the distribution of \(Z^2_3\), which follows a non-central \(\chi^2_6\) distribution\cite{Bachetti2022}. We then inverted the cumulative distribution function from the modeled probability density function for Equation~\ref{eq14} to determine the probability distribution of obtaining the measured power \(Z^2_3 = 63.32\) for various signal powers \(\hat{Z^2_3}\). We derived \(\hat{Z^2_3} = 58.40^{+16.5}_{-14.6}\), using \(f_n(Z^2_3 \mid \hat{Z^2_3}) = 0.5\) as the best estimate and \(f_n(Z^2_3 \mid \hat{Z^2_3}) = 0.159\) and \(0.841\) for the 1\(\sigma\) uncertainties, where \(f_n\) represents the cumulative distribution function. Consequently, the r.m.s. fractional amplitude of the 909 Hz signal in the optimal interval is estimated to be \(r = 27.20^{+3.60}_{-3.64}\%\).

\noindent\textbf{Signal from Different Detectors Onboard GECAM-B.}
Based on the optimal time interval and energy range determined from the combined data of GRD01, GRD04, and GRD05, we further evaluated whether the observed QPO could be an artifact of a single detector. Specifically, we assessed if the QPO is detectable across different detectors within the same time interval and if the QPO exhibits consistent energy dependence across these detectors.

To address these concerns, we first generated\cite{2021Natur.600..621C} dynamical power spectra using \(Z^2_1\) statistics, applying overlapping 160-ms intervals spaced at 7~ms within [24.241, 24.721]~s from $T_0$. As shown in Extended Data Fig. \ref{efig1} \textbf{a}, the dynamical power spectra from the three detectors revealed a consistent signal at 909~Hz within the same time interval (marked in the red box), suggesting that the QPO is unlikely to be an artifact of a single detector.

Next, to confirm that the QPO originates from the source rather than an instrumental effect, we examined the energy dependence of the QPO across different detectors. We segmented the data within the interval [24.401, 24.561]~s from \(T_0\) into 78 energy bands, each having a similar count rate to the 98--248~keV range used initially. Detailed energy ranges are listed in Extended Data Table \ref{etab1}, with a summary provided in Fig. \ref{fig3} \textbf{b} for the combined detectors and in Extended Data Fig. \ref{efig1} \textbf{b} for each detector individually.

Since the QPO is confirmed across different detectors in the optimal time interval, we calculated the r.m.s. fractional amplitude using \(Z^2_3\) statistics for each energy segment in [24.401, 24.561]~s from $T_0$. The results, shown in Extended Data Fig. \ref{efig1} \textbf{c}, demonstrate consistent energy dependence across different detectors. This uniformity supports the conclusion that the QPO was detected simultaneously by the detectors onboard GECAM-B from GRB~230307A, as observed from varying incident angles.

\noindent\textbf{Significance Estimation Based on Monte Carlo Simulations.}
Through the above steps, we determined the optimal time interval and energy range and validated that the observed QPO is not an artifact of instrumental effects. In the final step, we used Monte Carlo simulations to estimate the significance of the QPO. This was performed using binned light curves from the combined data of GRD01, GRD04, and GRD05. The process for generating synthetic light curves is outlined below:
\begin{enumerate}
\item The times of synthetic light curves were generated from the observation data by adding a random variate from a standard normal distribution with a standard deviation of 0.0001 (corresponding to the light curve bin size of 0.1~ms) to the observed time in seconds.
\item The counts of synthetic light curves were generated from the observation data by replacing the observed counts with a random variate from a Poisson distribution with an expectation of the observed value. To mitigate the ``red noise leak'' effect, we produced a surrogate data set 100 times the size of the observed one and randomly selected a subset with the same length as the observed data to obtain the counts of the artificial light curve\cite{2013MNRAS.433..907E}.
\end{enumerate}
In principle, simulating timestamps for synthetic light curves is not essential, as the intrinsic time resolution of the GECAM and Fermi/GBM detectors---on the order of microseconds---is far finer than the 0.1-ms bin size used in our analysis. Nonetheless, we incorporated timestamp perturbations to capture potential effects introduced by the binning process. In our construction of binned light curves from event data, each count is assigned to the centroid of its respective time bin, an approximation that may introduce subtle systematic biases. To model this, we added a random offset to each timestamp, drawn from a normal distribution with a standard deviation of 0.1 ms, corresponding to the bin width. This conservative treatment accounts for the inherent uncertainty in photon arrival times within each bin and yields a statistically robust estimate of any bias introduced by discretizing the light curve.

We generated \(1 \times 10^5\) sets of WWZ spectrograms from these synthetic light curves, using a time shift of 5~ms, a scale factor of \(2\pi \times 100\)~rad/s, and a frequency range of \(909 \pm 1/0.16\)~Hz, sampled at \(0.1/0.16\)~Hz resolution. Given the observed signal duration of 160~ms, we calculated the mean power of several 5-ms segments within the interval [24.401, 24.561]~s from $T_0$ to obtain the WWZ power distribution.

Theoretically, the WWZ power should follow an F-distribution with \(N_{\text{eff}} - 3\) and 2 degrees of freedom, where \(N_{\text{eff}}\) denotes the effective number of data points\cite{1996AJ....111..541F,1996AJ....111..555F}. However, because \(N_{\text{eff}}\) is not an integer, this exact F-distribution does not exist\cite{1996AJ....112.1709F}. To address this, we approximated the theoretical distribution by fitting the asymptotic distribution to the F-distribution using \texttt{scipy.stats}, which allowed us to reduce the computational complexity of replicating the observed extreme power. The probability density distributions for the observed and artificial WWZ powers are shown in Extended Data Fig. \ref{efig2} \textbf{b}. From the survival function of the asymptotic F-distribution, we calculated the single-trial chance probability of observing the 909 Hz signal to be approximately \(1.77 \times 10^{-8}\).

For comparison, we also assessed the significance of the 909~Hz signal within the interval [24.401, 24.561]~s from \(T_0\) in the 98--248~keV energy range using the Lomb-Scargle periodogram\cite{2018ApJS..236...16V}. The single-trial FAP obtained through the Baluev method\cite{2008MNRAS.385.1279B} was approximately \(6.15 \times 10^{-8}\).

\noindent\textbf{Consistency Check of the QPO.}
We conducted consistency checks across both the time and energy domains using observational data from GECAM-C and Fermi/GBM. We employed the WWZ algorithm to examine the frequency evolution in the binned light curves from these instruments, accounting for potential variations in the QPO's timescale, which could be affected by factors such as the incident angle or the detector area of each instrument. The wavelet transform was particularly effective in capturing the signal's characteristics over short timescales within the limited observation durations\cite{1996AJ....112.1709F}.

We extracted event data from GECAM-C within the 98--248~keV energy range using the combination of data from GRD01, GRD02 and GRD06. For Fermi/GBM, data are obtained from the detector na. As illustrated in Fig.~\ref{fig2}~\textbf{a}, the count rates from GECAM-C and Fermi/GBM are comparable. We applied a time shift of 5~ms and used a wavelet scale factor of $2\pi \times 100$~rad/s to generate WWZ spectrograms. We analysed 0.1-ms binned light curves from GECAM-B, GECAM-C, and Fermi/GBM within the time range [23.921, 25.041]~s from $T_0$, encompassing the optimal interval for the 909~Hz signal with a $300\%$ extension on either side. For each light curve, we sampled the WWZ power and calculated the fraction of the WWZ contour that maintained a duration of 160~ms for GECAM-B, which was approximately $97.76\%$ level. We then use the same level of WWZ powers to determine the contours from the spectrogram of GECAM-C and Fermi/GBM to ensure consistency across instruments. As shown in Fig. \ref{fig3} \textbf{a}, all instruments captured a similar signal peaking around 909~Hz simultaneously. Based on the time intervals defined by the WWZ contours at the 97.76\% level at 909~Hz, the signal's duration observed by GECAM-B is 160~ms. In comparison, GECAM-C recorded a similar duration of approximately 155~ms, while Fermi/GBM recorded a relatively longer duration of around 371~ms at the same threshold.

Additionally, for the time intervals during which the signal at 909~Hz was identified across different instruments, we measured the r.m.s. amplitudes across various energy ranges, as listed in Extended Data Table~\ref{etab1}. Each r.m.s. amplitude was independently calculated from the light curve (including background) within the corresponding energy range. The uncertainty in the r.m.s. amplitude corresponds to the 1$\sigma$ confidence interval, derived from the signal power using the measured $Z^2_1$ statistic at 909~Hz across different energy ranges within the identified time intervals for each instrument. The photon count rates and the corresponding r.m.s. amplitudes during the signal's time interval, across energy ranges and instruments, are presented in Fig.~\ref{fig3}~\textbf{b--c}. These measurements reveal a consistent energy-dependent structure of the signal across all three instruments, indicating that a similar QPO was simultaneously detected.

We also note that the duration of the QPO observed by Fermi/GBM is noticeably longer than that detected by the GECAM instruments. This discrepancy is likely due to the larger single-detector area of the Fermi/GBM instruments, which makes it easier to detect the QPO at the same count rate. Despite this, GECAM-B, with its higher flux at smaller incident angles, still achieved the highest r.m.s. fractional amplitude (Fig. \ref{fig3} \textbf{b-c}). The QPO-modulated counts (approximately equal to the product of the fractional amplitude and total counts) recorded by GECAM-C and Fermi/GBM are similar and align with the comparable incident angles of these detectors. These findings confirm that the signal detected across the three instruments originates from the same astrophysical source, most likely GRB~230307A.

\subsection{On the Quasi-Periodic Nature of the 909-Hz signal\\}
The QPO feature in the Fourier power density spectrum (PDS) can be characterized by a Lorentzian profile, which, when expressed as a probability density, is known as the Cauchy distribution, defined as follows\cite{Bachetti2022}:
\begin{equation}
P(f) = \frac{N}{\pi \Delta f_0}\frac{(\Delta f_0)^2}{(f - f_0)^2 + (\Delta f_0)^2},
\label{eq15}
\end{equation}
where $f$ is the frequency, $N$ is the amplitude, $f_0$ is the centroid frequency of the Lorentzian, and $\Delta f_0$ is the half width at half maximum of the Lorentzian.

The measured PDS with a normalized mean power of unity\cite{1975ApJS...29..285G}, compared with the expected model for the PDS, follows a similar form of probability distribution described in Equation~\ref{eq14}. Therefore, the likelihood of observing a measured power $P$ in a given frequency bin, given an expected model power $P_s$, is\cite{1975ApJS...29..285G,2023Natur.613..253C}:
\begin{equation}
\mathcal{L}(P\vert P_s)=e^{-(P+P_s)}\sum_{m=0}^\infty \frac{(P P_s)^m}{(m!)^2},
\label{eq16}
\end{equation}
where, following ref.\cite{2019ApJ...871...95M}, the infinite sum is approximated by truncating the series when the term being considered has a magnitude less than $10^{-20}$ of the cumulative sum. Using this approximation, we can calculate the likelihood of any given PDS model under the measured PDS and then use the Bayes factor between the white noise model and the white noise plus QPO model to determine if the measured PDS requires an additional QPO component. This approach assumes that only white noise, alongside unavoidable Poisson fluctuations, is present in the PDS\cite{2023Natur.613..253C}.

For comparison with the results of all short GRB samples in ref.\cite{2023Natur.613..253C}, we used a 100-ms segment length to obtain Groth's normalized\cite{1975ApJS...29..285G} PDS within the interval [24.401, 24.561]~s from $T_0$ in the 98--248~keV range. The optimal time segment was selected based on the signal with the highest power at 910~Hz. The spectrogram is presented in Extended Data Fig. \ref{efig3} \textbf{a}, with the optimal segment identified as [24.451, 24.551]~s from $T_0$.

We then used the likelihood function described in Equation~\ref{eq16} to fit the PDS with both the white noise model and the white noise plus QPO model. In this context, the Lorentzian representing the QPO component is simplified as $A=\frac{N}{\pi \Delta f_0}$, where $A$ represents the power density at $f_0$. The best-fit parameters were derived from the posterior distributions obtained using the nested sampling algorithm package \texttt{MULTINEST}\cite{2009MNRAS.398.1601F}. We chose \texttt{MULTINEST} over Markov chain Monte Carlo simulations for Bayesian inference because it facilitates obtaining Bayesian evidence without needing to set initial prior parameter values. We maintained the prior interval of model parameters as in ref.\cite{2023Natur.613..253C}, except for a flat prior on the central frequency parameter and a smaller lower limit $\log_{10} (0.1~\text{Hz})$ for the width parameter. This adjustment was necessary because the QPO signal in GRB~230307A appears to be periodic, and the original parameter settings risked placing the best-fit results on the boundary of the pairwise marginal posterior distributions.

The observed PDS and the comparison of the Bayes factor distributions are shown in Extended Data Fig. \ref{efig3} \textbf{c, d}. The PDS reveals no notable red noise excess in the 10--500~Hz range, suggesting that the chance probability derived from theoretical distributions, which account solely for Poisson noise, remains valid. Moreover, despite the signal's high coherence (with a quality factor $Q = \frac{f_0}{2\Delta f_0} = 476$, for example, refs.\cite{2000MNRAS.318..361N,2012ARA&A..50..609W,Bachetti2022}) and an extreme power excess at a single frequency bin, the Bayes factor comparing the white noise plus QPO model to the white noise model still achieved a remarkable value, similar to those found in GRB~931101B and GRB~970711.
This indicates that the observed PDS cannot be described by the white noise model alone, strongly supporting the presence of a QPO component at~909 Hz. Detailed fitting results for different models are presented in Extended Data Tab. \ref{etab2}.

\subsection{Temporal Coincidence Between the 909-Hz Signal and HLE\\}
The detection of the 909-Hz signal at a later stage in GRB~230307A prompted us to investigate its potential connection to a specific phase following the merger. To explore this, we used time-resolved spectral fitting results from ref.\cite{2023arXiv230705689S} to analyze the spectral flux evolution of GRB~230307A, as shown in Extended Data Fig. \ref{efig4}. The spectral flux was calculated within the 98--248~keV range, which is optimal for detecting the 909-Hz signal.

Extended Data Fig. \ref{efig4} demonstrates that the 909-Hz signal coincides with a transition phase associated with the ``curvature effect''\cite{2000ApJ...541L..51K,2004ApJ...614..284D}. This transition marks the onset of HLE from regions outside the $1/\Gamma$ jet cone, which begins to reach the observer's line of sight and dominate the gamma-ray flux, causing a steeper decay slope. The concurrent occurrence of the ``curvature effect'' and the 909-Hz signal suggests that the periodicity was detected at a pivotal epoch, where the emission shifted from being dominated by the prompt emission to being influenced by HLE.

\subsection{Blind search for periodicity in GRB~211211A\\}

Another bright GRB with a duration of approximately one minute, GRB~211211A, exhibits many observational similarities to GRB~230307A. This motivated us to conduct a blind search for periodicity in GRB~211211A. Data from Fermi/GBM were extracted using a combination of two NaI detectors, n2 and na, which had the smallest incident angles relative to the source. For the combined event data of GRB~211211A, the Bayesian block duration was determined to be [-0.016, 81.901]~s from the Fermi/GBM trigger time (hereafter $T_{0,{\rm 211211A}}$)\cite{2021GCN.31201....1F}, with MVT of 5~ms. For 100-ms segments with SNR $\geq$ 10 in GRB~211211A, the average count rate was estimated to be approximately 5,200 photons per second. From these, we selected roughly 19\% as the fraction used to generate subsets from each segment, resulting in a total of 462,852 subsets derived from 8,935 time segments, following the same segmentation strategy applied to GRB~230307A. The background search yielded a false alarm threshold of $R_{\rm false} = 15.85$ for GRB~211211A. The candidate frequencies and corresponding power values are presented in Extended Data Fig.~\ref{efig5}.

As shown in Extended Data Fig.~\ref{efig5} \textbf{a--c}, the maximum candidate power was observed at 546~Hz, with a Rayleigh power of $R = 37.97$. By tracing the time intervals of the subsets producing this extreme power, we identified a signal occurring between 14.880 and 15.075~s from $T_{0,{\rm 211211A}}$. This signal exhibited candidate powers exceeding $R_{\rm false} = 15.85$ across a broad frequency range of 500--549~Hz, suggesting it could either be an potential QPO or a broadband noise. As presented in Extended Data Fig.~\ref{efig5} \textbf{d--e}, we computed the FAPs for each trial frequency. Within the 500--549~Hz range, the minimum probability was found at 500~Hz, with a value of $\approx 31.09\%$ after accounting for 2,001 trial frequencies. The representative FAP for this signal, calculated as the product of all probabilities in the 500--549~Hz range, was $\approx 31.09\%$, indicating that the signal is most likely due to noise.

Although the representative FAP is high, the single-trial extreme power of $R = 37.97$ remains unusually large for an exponential distribution with an expected value of 1, corresponding to a chance probability of $\approx 3.23 \times 10^{-17}$. Therefore, we further examined this signal using the same procedure applied to GRB~230307A to determine its optimal time and energy range. As shown in Extended Data Fig.~\ref{efig6}, we identified a very bright spike within the [14.880, 15.075]~s range from $T_{0,{\rm 211211A}}$ in the 5--20~keV band. This spike lasted approximately 1~ms, generating strong red noise exceeding 500~Hz and resulting in the detection of extreme Rayleigh power within the 500--549~Hz range. Additionally, this spike was detected only by detector na, which had a larger incident angle than n2. The discrepancy between the detectors suggests that this signal is not associated with GRB~211211A. The PDS in Extended Data Fig.~\ref{efig6} further confirmed that the extreme power detection in Extended Data Fig.~\ref{efig5} was due to red noise. However, our statistical model had already rejected such signals due to strong red noise based on the derived FAP.

Additionally, as shown in Extended Data Fig.~\ref{efig5} \textbf{e}, another signal at 935~Hz exhibited a relatively low FAP, surpassing the $2\sigma$ significance level, and showed candidate power above the threshold of $R_{\rm false} = 15.85$ in Extended Data Fig.~\ref{efig5} \textbf{b}. This signal was detected within the [50.914, 51.029]~s range from $T_{0,{\rm 211211A}}$, with 8 candidate powers exceeding $R_{\rm false}$ at both 935~Hz and 936~Hz. The representative FAP for this signal was calculated to be approximately 1.09\%. Although the 935~Hz frequency is close to the magnetar spin frequency of 909~Hz in GRB~230307A, the relatively large FAP and the lack of constraint on the transition time due to the ``curvature effect'' in GRB~211211A\cite{2022Natur.612..232Y} make this signal inconclusive. This low-significance detection suggests that identifying QPOs in GRBs may require a high photon count during the prompt emission phase, as well as a Poynting-flux-dominated outflow capable of generating mini-jets that act as hot spots.

\clearpage

\section*{Data Availability}

The processed data presented in the tables and figures of the paper are available upon reasonable request. The authors also note that some of the data used in this study are accessible through the NASA/Fermi Data Archive, the GECAM Data Archive, or GCN circulars. For GECAM data of GRB~230307A, which are currently be under the protection period and not publicly accessible, researchers or organizations may submit a reasonable request via email to obtain and utilize the proprietary data. For data access inquiries, please contact S.-L. X (xiongsl@ihep.ac.cn).

\section*{Code Availability}
Upon reasonable requests, the code (mostly in Python) used to produce the results and figures will be provided.

\clearpage


\begin{thebibliography}{10}
\expandafter\ifx\csname url\endcsname\relax
  \def\url#1{\texttt{#1}}\fi
\expandafter\ifx\csname urlprefix\endcsname\relax\def\urlprefix{URL }\fi
\providecommand{\bibinfo}[2]{#2}
\providecommand{\eprint}[2][]{\url{#2}}

\bibitem{2014ARA&A..52...43B}
\bibinfo{author}{{Berger}, E.}
\newblock \bibinfo{title}{{Short-Duration Gamma-Ray Bursts}}.
\newblock \emph{\bibinfo{journal}{\araa}} \textbf{\bibinfo{volume}{52}}, \bibinfo{pages}{43--105} (\bibinfo{year}{2014}).

\bibitem{Zhang_2018}
\bibinfo{author}{Zhang, B.}
\newblock \emph{\bibinfo{title}{The Physics of Gamma-Ray Bursts}} (\bibinfo{publisher}{Cambridge Univ. Press}, \bibinfo{year}{2018}).

\bibitem{2017PhRvL.119p1101A}
\bibinfo{author}{{Abbott}, B.~P.} \emph{et~al.}
\newblock \bibinfo{title}{{GW170817: Observation of Gravitational Waves from a Binary Neutron Star Inspiral}}.
\newblock \emph{\bibinfo{journal}{\prl}} \textbf{\bibinfo{volume}{119}}, \bibinfo{pages}{161101} (\bibinfo{year}{2017}).

\bibitem{1997ApJ...490..633K}
\bibinfo{author}{{Katz}, J.~I.}
\newblock \bibinfo{title}{{Yet Another Model of Gamma-Ray Bursts}}.
\newblock \emph{\bibinfo{journal}{\apj}} \textbf{\bibinfo{volume}{490}}, \bibinfo{pages}{633--641} (\bibinfo{year}{1997}).

\bibitem{2006A&A...454...11R}
\bibinfo{author}{{Reynoso}, M.~M.}, \bibinfo{author}{{Romero}, G.~E.} \& \bibinfo{author}{{Sampayo}, O.~A.}
\newblock \bibinfo{title}{{Precession of neutrino-cooled accretion disks in gamma-ray burst engines}}.
\newblock \emph{\bibinfo{journal}{\aap}} \textbf{\bibinfo{volume}{454}}, \bibinfo{pages}{11--16} (\bibinfo{year}{2006}).

\bibitem{2010A&A...516A..16L}
\bibinfo{author}{{Liu}, T.} \emph{et~al.}
\newblock \bibinfo{title}{{Jet precession driven by neutrino-cooled disk for gamma-ray bursts}}.
\newblock \emph{\bibinfo{journal}{\aap}} \textbf{\bibinfo{volume}{516}}, \bibinfo{pages}{A16} (\bibinfo{year}{2010}).

\bibitem{2005PhRvL..94t1101S}
\bibinfo{author}{{Shibata}, M.}
\newblock \bibinfo{title}{{Constraining Nuclear Equations of State Using Gravitational Waves from Hypermassive Neutron Stars}}.
\newblock \emph{\bibinfo{journal}{\prl}} \textbf{\bibinfo{volume}{94}}, \bibinfo{pages}{201101} (\bibinfo{year}{2005}).

\bibitem{2014PhRvL.113i1104T}
\bibinfo{author}{{Takami}, K.}, \bibinfo{author}{{Rezzolla}, L.} \& \bibinfo{author}{{Baiotti}, L.}
\newblock \bibinfo{title}{{Constraining the Equation of State of Neutron Stars from Binary Mergers}}.
\newblock \emph{\bibinfo{journal}{\prl}} \textbf{\bibinfo{volume}{113}}, \bibinfo{pages}{091104} (\bibinfo{year}{2014}).

\bibitem{2021GReGr..53...59S}
\bibinfo{author}{{Sarin}, N.} \& \bibinfo{author}{{Lasky}, P.~D.}
\newblock \bibinfo{title}{{The evolution of binary neutron star post-merger remnants: a review}}.
\newblock \emph{\bibinfo{journal}{Gen. Relativ. Gravit.}} \textbf{\bibinfo{volume}{53}}, \bibinfo{pages}{59} (\bibinfo{year}{2021}).

\bibitem{2013PhRvD..87h4053S}
\bibinfo{author}{{Stone}, N.}, \bibinfo{author}{{Loeb}, A.} \& \bibinfo{author}{{Berger}, E.}
\newblock \bibinfo{title}{{Pulsations in short gamma ray bursts from black hole-neutron star mergers}}.
\newblock \emph{\bibinfo{journal}{\prd}} \textbf{\bibinfo{volume}{87}}, \bibinfo{pages}{084053} (\bibinfo{year}{2013}).

\bibitem{Bachetti2022}
\bibinfo{author}{Bachetti, M.} \& \bibinfo{author}{Huppenkothen, D.}
\newblock \emph{\bibinfo{title}{{\rm in} Handbook of X-ray and Gamma-ray Astrophysics}} (\bibinfo{editor}{eds Bambi, C. \& Santangelo, A.}), \bibinfo{pages}{1--47} (\bibinfo{publisher}{Springer Nature Singapore}, \bibinfo{year}{2022}).

\bibitem{2022ApJS..259...32H}
\bibinfo{author}{{H{\"u}bner}, M.}, \bibinfo{author}{{Huppenkothen}, D.}, \bibinfo{author}{{Lasky}, P.~D.} \& \bibinfo{author}{{Inglis}, A.~R.}
\newblock \bibinfo{title}{{Pitfalls of Periodograms: The Nonstationarity Bias in the Analysis of Quasiperiodic Oscillations}}.
\newblock \emph{\bibinfo{journal}{\\apjs}} \textbf{\bibinfo{volume}{259}}, \bibinfo{pages}{32} (\bibinfo{year}{2022}).

\bibitem{2023Natur.613..253C}
\bibinfo{author}{{Chirenti}, C.}, \bibinfo{author}{{Dichiara}, S.}, \bibinfo{author}{{Lien}, A.}, \bibinfo{author}{{Miller}, M.~C.} \& \bibinfo{author}{{Preece}, R.}
\newblock \bibinfo{title}{{Kilohertz quasiperiodic oscillations in short gamma-ray bursts}}.
\newblock \emph{\bibinfo{journal}{\\nat}} \textbf{\bibinfo{volume}{613}}, \bibinfo{pages}{253--256} (\bibinfo{year}{2023}).

\bibitem{2024Natur.626..737L}
\bibinfo{author}{{Levan}, A.~J.} \emph{et~al.}
\newblock \bibinfo{title}{{Heavy-element production in a compact object merger observed by JWST}}.
\newblock \emph{\bibinfo{journal}{\nat}} \textbf{\bibinfo{volume}{626}}, \bibinfo{pages}{737--741} (\bibinfo{year}{2024}).

\bibitem{2024Natur.626..742Y}
\bibinfo{author}{{Yang}, Y.-H.} \emph{et~al.}
\newblock \bibinfo{title}{{A lanthanide-rich kilonova in the aftermath of a long gamma-ray burst}}.
\newblock \emph{\bibinfo{journal}{\nat}} \textbf{\bibinfo{volume}{626}}, \bibinfo{pages}{742--745} (\bibinfo{year}{2024}).

\bibitem{2023arXiv230705689S}
\bibinfo{author}{Sun, H.} \emph{et~al.}
\newblock \bibinfo{title}{Magnetar emergence in a peculiar gamma-ray burst from a compact star merger}.
\newblock \emph{\bibinfo{journal}{Natl Sci. Rev.}} \bibinfo{pages}{nwae401} (\bibinfo{year}{2024}).

\bibitem{Li2021RDTM}
\bibinfo{author}{Li, X.~Q.} \emph{et~al.}
\newblock \bibinfo{title}{{The technology for detection of gamma-ray burst with GECAM satellite}}.
\newblock \emph{\bibinfo{journal}{Radiat. Detect. Technol. Methods}} \textbf{\bibinfo{volume}{6}}, \bibinfo{pages}{12--25} (\bibinfo{year}{2021}).

\bibitem{2023NIMPA105668586Z}
\bibinfo{author}{{Zhang}, D.} \emph{et~al.}
\newblock \bibinfo{title}{{The performance of SiPM-based gamma-ray detector (GRD) of GECAM-C}}.
\newblock \emph{\bibinfo{journal}{Nucl. Instrum. Methods Phys. Res. Sect. A}} \textbf{\bibinfo{volume}{1056}}, \bibinfo{pages}{168586} (\bibinfo{year}{2023}).

\bibitem{2009ApJ...702..791M}
\bibinfo{author}{{Meegan}, C.} \emph{et~al.}
\newblock \bibinfo{title}{{The Fermi Gamma-ray Burst Monitor}}.
\newblock \emph{\bibinfo{journal}{\apj}} \textbf{\bibinfo{volume}{702}}, \bibinfo{pages}{791--804} (\bibinfo{year}{2009}).

\bibitem{2023GCN.33551....1D}
\bibinfo{author}{{Dalessi}, S.} \& \bibinfo{author}{{Fermi GBM Team}}.
\newblock \bibinfo{title}{{GRB 230307A: Bad Time Intervals for Fermi GBM data}}.
\newblock \emph{\bibinfo{journal}{GRB Coordinates Network}} \textbf{\bibinfo{volume}{33551}}, \bibinfo{pages}{1} (\bibinfo{year}{2023}).

\bibitem{2023GCN.33406....1X}
\bibinfo{author}{{Xiong}, S.}, \bibinfo{author}{{Wang}, C.}, \bibinfo{author}{{Huang}, Y.} \& \bibinfo{author}{{Gecam Team}}.
\newblock \bibinfo{title}{{GRB 230307A: GECAM detection of an extremely bright burst}}.
\newblock \emph{\bibinfo{journal}{GRB Coordinates Network}} \textbf{\bibinfo{volume}{33406}}, \bibinfo{pages}{1} (\bibinfo{year}{2023}).

\bibitem{1928PCPS...24..180F}
\bibinfo{author}{{Fisher}, R.~A.} \& \bibinfo{author}{{Tippett}, L.~H.~C.}
\newblock \bibinfo{title}{{Limiting forms of the frequency distribution of the largest or smallest member of a sample}}.
\newblock \emph{\bibinfo{journal}{Math. Proc. of the Camb. Philos. Soc.}} \textbf{\bibinfo{volume}{24}}, \bibinfo{pages}{180} (\bibinfo{year}{1928}).

\bibitem{mood1950introduction}
\bibinfo{author}{Mood, A.~M.}
\newblock \emph{\bibinfo{title}{Introduction to the theory of statistics}} (\bibinfo{publisher}{McGraw-Hill}, \bibinfo{year}{1950}).

\bibitem{1983A&A...128..245B}
\bibinfo{author}{{Buccheri}, R.} \emph{et~al.}
\newblock \bibinfo{title}{{Search for pulsed {\ensuremath{\gamma}}-ray emission from radio pulsars in the COS-B data.}}
\newblock \emph{\bibinfo{journal}{\aap}} \textbf{\bibinfo{volume}{128}}, \bibinfo{pages}{245--251} (\bibinfo{year}{1983}).

\bibitem{2021Natur.600..621C}
\bibinfo{author}{{Castro-Tirado}, A.~J.} \emph{et~al.}
\newblock \bibinfo{title}{{Very-high-frequency oscillations in the main peak of a magnetar giant flare}}.
\newblock \emph{\bibinfo{journal}{\nat}} \textbf{\bibinfo{volume}{600}}, \bibinfo{pages}{621--624} (\bibinfo{year}{2021}).

\bibitem{2012ARA&A..50..609W}
\bibinfo{author}{{Watts}, A.~L.}
\newblock \bibinfo{title}{{Thermonuclear Burst Oscillations}}.
\newblock \emph{\bibinfo{journal}{\araa}} \textbf{\bibinfo{volume}{50}}, \bibinfo{pages}{609--640} (\bibinfo{year}{2012}).

\bibitem{1989A&A...221..180D}
\bibinfo{author}{{de Jager}, O.~C.}, \bibinfo{author}{{Raubenheimer}, B.~C.} \& \bibinfo{author}{{Swanepoel}, J.~W.~H.}
\newblock \bibinfo{title}{{A powerful test for weak periodic signals with unknown light curve shape in sparse data.}}
\newblock \emph{\bibinfo{journal}{\aap}} \textbf{\bibinfo{volume}{221}}, \bibinfo{pages}{180--190} (\bibinfo{year}{1989}).

\bibitem{2005ApJ...634..547W}
\bibinfo{author}{{Watts}, A.~L.}, \bibinfo{author}{{Strohmayer}, T.~E.} \& \bibinfo{author}{{Markwardt}, C.~B.}
\newblock \bibinfo{title}{{Analysis of Variability in the Burst Oscillations of the Accreting Millisecond Pulsar XTE J1814-338}}.
\newblock \emph{\bibinfo{journal}{\apj}} \textbf{\bibinfo{volume}{634}}, \bibinfo{pages}{547--564} (\bibinfo{year}{2005}).

\bibitem{1996AJ....112.1709F}
\bibinfo{author}{{Foster}, G.}
\newblock \bibinfo{title}{{Wavelets for period analysis of unevenly sampled time series}}.
\newblock \emph{\bibinfo{journal}{\aj}} \textbf{\bibinfo{volume}{112}}, \bibinfo{pages}{1709--1729} (\bibinfo{year}{1996}).

\bibitem{1975ApJS...29..285G}
\bibinfo{author}{{Groth}, E.~J.}
\newblock \bibinfo{title}{{Probability distributions related to power spectra.}}
\newblock \emph{\bibinfo{journal}{\apjs}} \textbf{\bibinfo{volume}{29}}, \bibinfo{pages}{285--302} (\bibinfo{year}{1975}).

\bibitem{2000MNRAS.318..361N}
\bibinfo{author}{{Nowak}, M.~A.}
\newblock \bibinfo{title}{{Are there three peaks in the power spectra of GX 339-4 and Cyg X-1?}}
\newblock \emph{\bibinfo{journal}{\mnras}} \textbf{\bibinfo{volume}{318}}, \bibinfo{pages}{361--367} (\bibinfo{year}{2000}).

\bibitem{2000ApJ...541L..51K}
\bibinfo{author}{{Kumar}, P.} \& \bibinfo{author}{{Panaitescu}, A.}
\newblock \bibinfo{title}{{Afterglow Emission from Naked Gamma-Ray Bursts}}.
\newblock \emph{\bibinfo{journal}{\apjl}} \textbf{\bibinfo{volume}{541}}, \bibinfo{pages}{L51--L54} (\bibinfo{year}{2000}).

\bibitem{2004ApJ...614..284D}
\bibinfo{author}{{Dermer}, C.~D.}
\newblock \bibinfo{title}{{Curvature Effects in Gamma-Ray Burst Colliding Shells}}.
\newblock \emph{\bibinfo{journal}{\apj}} \textbf{\bibinfo{volume}{614}}, \bibinfo{pages}{284--292} (\bibinfo{year}{2004}).

\bibitem{2000ARA&A..38..717V}
\bibinfo{author}{{van der Klis}, M.}
\newblock \bibinfo{title}{{Millisecond Oscillations in X-ray Binaries}}.
\newblock \emph{\bibinfo{journal}{\araa}} \textbf{\bibinfo{volume}{38}}, \bibinfo{pages}{717--760} (\bibinfo{year}{2000}).

\bibitem{2011ApJ...726...90Z}
\bibinfo{author}{{Zhang}, B.} \& \bibinfo{author}{{Yan}, H.}
\newblock \bibinfo{title}{{The Internal-collision-induced Magnetic Reconnection and Turbulence (ICMART) Model of Gamma-ray Bursts}}.
\newblock \emph{\bibinfo{journal}{\apj}} \textbf{\bibinfo{volume}{726}}, \bibinfo{pages}{90} (\bibinfo{year}{2011}).

\bibitem{2023arXiv231007205Y}
\bibinfo{author}{{Yi}, S.~X.} \emph{et~al.}
\newblock \bibinfo{title}{{Evidence of minijet emission in a large emission zone from a magnetically dominated gamma-ray burst jet}}.
\newblock \emph{\bibinfo{journal}{\apj}} \textbf{\bibinfo{volume}{985}}, \bibinfo{pages}{239} (\bibinfo{year}{2025}).

\bibitem{2022Natur.612..223R}
\bibinfo{author}{{Rastinejad}, J.~C.} \emph{et~al.}
\newblock \bibinfo{title}{{A kilonova following a long-duration gamma-ray burst at 350 Mpc}}.
\newblock \emph{\bibinfo{journal}{\nat}} \textbf{\bibinfo{volume}{612}}, \bibinfo{pages}{223--227} (\bibinfo{year}{2022}).

\bibitem{2022Natur.612..228T}
\bibinfo{author}{{Troja}, E.} \emph{et~al.}
\newblock \bibinfo{title}{{A nearby long gamma-ray burst from a merger of compact objects}}.
\newblock \emph{\bibinfo{journal}{\nat}} \textbf{\bibinfo{volume}{612}}, \bibinfo{pages}{228--231} (\bibinfo{year}{2022}).

\bibitem{2022Natur.612..232Y}
\bibinfo{author}{{Yang}, J.} \emph{et~al.}
\newblock \bibinfo{title}{{A long-duration gamma-ray burst with a peculiar origin}}.
\newblock \emph{\bibinfo{journal}{\nat}} \textbf{\bibinfo{volume}{612}}, \bibinfo{pages}{232--235} (\bibinfo{year}{2022}).

\bibitem{2024ApJ...970....6X}
\bibinfo{author}{{Xiao}, S.} \emph{et~al.}
\newblock \bibinfo{title}{{The Peculiar Precursor of a Gamma-Ray Burst from a Binary Merger Involving a Magnetar}}.
\newblock \emph{\bibinfo{journal}{\apj}} \textbf{\bibinfo{volume}{970}}, \bibinfo{pages}{6} (\bibinfo{year}{2024}).

\bibitem{2024ApJ...967...26C}
\bibinfo{author}{{Chirenti}, C.}, \bibinfo{author}{{Dichiara}, S.}, \bibinfo{author}{{Lien}, A.} \& \bibinfo{author}{{Miller}, M.~C.}
\newblock \bibinfo{title}{{Evidence of a Strong 19.5 Hz Flux Oscillation in Swift BAT and Fermi GBM Gamma-Ray Data from GRB 211211A}}.
\newblock \emph{\bibinfo{journal}{\apj}} \textbf{\bibinfo{volume}{967}}, \bibinfo{pages}{26} (\bibinfo{year}{2024}).

\bibitem{2025JHEAp..45..325Z}
\bibinfo{author}{{Zhang}, B.}
\newblock \bibinfo{title}{{On the duration of gamma-ray bursts}}.
\newblock \emph{\bibinfo{journal}{J. High Energy Astrophys.}} \textbf{\bibinfo{volume}{45}}, \bibinfo{pages}{325--332} (\bibinfo{year}{2025}).

\bibitem{1996AJ....111..541F}
\bibinfo{author}{{Foster}, G.}
\newblock \bibinfo{title}{{Time Series Analysis by Projection. I. Statistical Properties of Fourier Analysis}}.
\newblock \emph{\bibinfo{journal}{\aj}} \textbf{\bibinfo{volume}{111}}, \bibinfo{pages}{541} (\bibinfo{year}{1996}).

\bibitem{2001A&A...369..694S}
\bibinfo{author}{{Spruit}, H.~C.}, \bibinfo{author}{{Daigne}, F.} \& \bibinfo{author}{{Drenkhahn}, G.}
\newblock \bibinfo{title}{{Large scale magnetic fields and their dissipation in GRB fireballs}}.
\newblock \emph{\bibinfo{journal}{\aap}} \textbf{\bibinfo{volume}{369}}, \bibinfo{pages}{694--705} (\bibinfo{year}{2001}).

\bibitem{2025MNRAS.tmp..154Y}
\bibinfo{author}{{Yang}, X.}, \bibinfo{author}{{L{\"u}}, H.-J.}, \bibinfo{author}{{Rice}, J.} \& \bibinfo{author}{{Liang}, E.-W.}
\newblock \bibinfo{title}{{Discovery of high-frequency quasi-periodic oscillation in short-duration gamma-ray bursts}}.
\newblock \emph{\bibinfo{journal}{\mnras}}  (\bibinfo{year}{2025}).

\bibitem{2011arXiv1111.0514W}
\bibinfo{author}{{Watts}, A.~L.}
\newblock \bibinfo{title}{{Neutron starquakes and the dynamic crust}}.
\newblock \emph{\bibinfo{journal}{arXiv e-prints}} \bibinfo{pages}{arXiv:1111.0514} (\bibinfo{year}{2011}).

\bibitem{2015ApJ...811...93G}
\bibinfo{author}{{Golkhou}, V.~Z.}, \bibinfo{author}{{Butler}, N.~R.} \& \bibinfo{author}{{Littlejohns}, O.~M.}
\newblock \bibinfo{title}{{The Energy Dependence of GRB Minimum Variability Timescales}}.
\newblock \emph{\bibinfo{journal}{\apj}} \textbf{\bibinfo{volume}{811}}, \bibinfo{pages}{93} (\bibinfo{year}{2015}).

\bibitem{2023A&A...671A.112C}
\bibinfo{author}{{Camisasca}, A.~E.} \emph{et~al.}
\newblock \bibinfo{title}{{GRB minimum variability timescale with Insight-HXMT and Swift. Implications for progenitor models, dissipation physics, and GRB classifications}}.
\newblock \emph{\bibinfo{journal}{\aap}} \textbf{\bibinfo{volume}{671}}, \bibinfo{pages}{A112} (\bibinfo{year}{2023}).

\bibitem{2013ApJ...764..167S}
\bibinfo{author}{{Scargle}, J.~D.}, \bibinfo{author}{{Norris}, J.~P.}, \bibinfo{author}{{Jackson}, B.} \& \bibinfo{author}{{Chiang}, J.}
\newblock \bibinfo{title}{{Studies in Astronomical Time Series Analysis. VI. Bayesian Block Representations}}.
\newblock \emph{\bibinfo{journal}{\apj}} \textbf{\bibinfo{volume}{764}}, \bibinfo{pages}{167} (\bibinfo{year}{2013}).

\bibitem{2018ApJS..236...17V}
\bibinfo{author}{{Vianello}, G.}
\newblock \bibinfo{title}{{The Significance of an Excess in a Counting Experiment: Assessing the Impact of Systematic Uncertainties and the Case with a Gaussian Background}}.
\newblock \emph{\bibinfo{journal}{\apjs}} \textbf{\bibinfo{volume}{236}}, \bibinfo{pages}{17} (\bibinfo{year}{2018}).

\bibitem{10.1111/j.2517-6161.1975.tb01550.x}
\bibinfo{author}{Mardia, K.~V.}
\newblock \bibinfo{title}{{Statistics of Directional Data}}.
\newblock \emph{\bibinfo{journal}{J. R. Stat. Soc. Ser. B}} \textbf{\bibinfo{volume}{37}}, \bibinfo{pages}{349--371} (\bibinfo{year}{2018}).

\bibitem{2002ApJ...576..932K}
\bibinfo{author}{{Kruger}, A.~T.}, \bibinfo{author}{{Loredo}, T.~J.} \& \bibinfo{author}{{Wasserman}, I.}
\newblock \bibinfo{title}{{Search for High-Frequency Periodicities in Time-tagged Event Data from Gamma-Ray Bursts and Soft Gamma Repeaters}}.
\newblock \emph{\bibinfo{journal}{\apj}} \textbf{\bibinfo{volume}{576}}, \bibinfo{pages}{932--941} (\bibinfo{year}{2002}).

\bibitem{1994ApJ...435..362V}
\bibinfo{author}{{Vaughan}, B.~A.} \emph{et~al.}
\newblock \bibinfo{title}{{Searches for Millisecond Pulsations in Low-Mass X-Ray Binaries. II.}}
\newblock \emph{\bibinfo{journal}{\apj}} \textbf{\bibinfo{volume}{435}}, \bibinfo{pages}{362} (\bibinfo{year}{1994}).

\bibitem{2013MNRAS.433..907E}
\bibinfo{author}{{Emmanoulopoulos}, D.}, \bibinfo{author}{{McHardy}, I.~M.} \& \bibinfo{author}{{Papadakis}, I.~E.}
\newblock \bibinfo{title}{{Generating artificial light curves: revisited and updated}}.
\newblock \emph{\bibinfo{journal}{\mnras}} \textbf{\bibinfo{volume}{433}}, \bibinfo{pages}{907--927} (\bibinfo{year}{2013}).

\bibitem{1996AJ....111..555F}
\bibinfo{author}{{Foster}, G.}
\newblock \bibinfo{title}{{Time Series Analysis by Projection. II. Tensor Methods for Time Series Analysis}}.
\newblock \emph{\bibinfo{journal}{\aj}} \textbf{\bibinfo{volume}{111}}, \bibinfo{pages}{555} (\bibinfo{year}{1996}).

\bibitem{2018ApJS..236...16V}
\bibinfo{author}{{VanderPlas}, J.~T.}
\newblock \bibinfo{title}{{Understanding the Lomb-Scargle Periodogram}}.
\newblock \emph{\bibinfo{journal}{\apjs}} \textbf{\bibinfo{volume}{236}}, \bibinfo{pages}{16} (\bibinfo{year}{2018}).

\bibitem{2008MNRAS.385.1279B}
\bibinfo{author}{{Baluev}, R.~V.}
\newblock \bibinfo{title}{{Assessing the statistical significance of periodogram peaks}}.
\newblock \emph{\bibinfo{journal}{\mnras}} \textbf{\bibinfo{volume}{385}}, \bibinfo{pages}{1279--1285} (\bibinfo{year}{2008}).

\bibitem{2019ApJ...871...95M}
\bibinfo{author}{{Miller}, M.~C.}, \bibinfo{author}{{Chirenti}, C.} \& \bibinfo{author}{{Strohmayer}, T.~E.}
\newblock \bibinfo{title}{{On the Persistence of QPOs during the SGR 1806-20 Giant Flare}}.
\newblock \emph{\bibinfo{journal}{\apj}} \textbf{\bibinfo{volume}{871}}, \bibinfo{pages}{95} (\bibinfo{year}{2019}).

\bibitem{2009MNRAS.398.1601F}
\bibinfo{author}{{Feroz}, F.}, \bibinfo{author}{{Hobson}, M.~P.} \& \bibinfo{author}{{Bridges}, M.}
\newblock \bibinfo{title}{{MULTINEST: an efficient and robust Bayesian inference tool for cosmology and particle physics}}.
\newblock \emph{\bibinfo{journal}{\mnras}} \textbf{\bibinfo{volume}{398}}, \bibinfo{pages}{1601--1614} (\bibinfo{year}{2009}).

\bibitem{2021GCN.31201....1F}
\bibinfo{author}{{Fermi GBM Team}}.
\newblock \bibinfo{title}{{GRB 211211A: Fermi GBM Final Real-time Localization}}.
\newblock \emph{\bibinfo{journal}{GRB Coordinates Network}} \textbf{\bibinfo{volume}{31201}}, \bibinfo{pages}{1} (\bibinfo{year}{2021}).

\end{thebibliography}

\clearpage

\begin{addendum}

\item[Acknowledgments] 
We thank Zhen-Yu Yan, Rong-Feng Shen, and He Gao for helpful comments. We are grateful to the GECAM team for the development and operation of the GECAM mission. The GECAM mission is supported by the Strategic Priority Research Program on Space Science of the Chinese Academy of Sciences.
This work was supported by the National Key Research and Development Programs of China (grants 2022YFF0711404 and 2022SKA0130102 to B.-B.Z., 2021YFA0718500 to S.-L.X., C.-W.W., W.-J.T., and S.-N.Z.), the National SKA Program of China (grant 2022SKA0130100 to B.-B.Z.), the Strategic Priority Research Program of the Chinese Academy of Sciences (grants XDA15360102, XDA15360300 and XDB0550300 to S.-L.X., C.-W.W., W.-J.T., and S.-N.Z.), and the National Natural Science Foundation of China (grants 11833003, U2038105 and 12121003 to B.-B.Z., 13001106 to J.Y., 12273042 and 12494572, to S.-L.X., C.-W.W., and W.-J.T., 12333007 to S.-N.Z.). B.-B.Z acknowledge support by the science research grants from the China Manned Space Program (grant CMS-CSST-2021-B11), the Fundamental Research Funds for the Central Universities, and the Program for Innovative Talents and Entrepreneurs in Jiangsu.
This work was performed on an HPC server equipped with two Intel Xeon Gold 6248 processors at Nanjing University. We acknowledge IT support from the computer lab of the School of Astronomy and Space Science at Nanjing University.

\item[Author Contributions]
R.-C.C. and B.-B.Z. initiated the study. B.-B.Z., B.Z. and R.-C.C. coordinated the scientific investigations of the event. R.-C.C., J.Y. and Y.-H.I.Y. performed the published data acquisition. R.-C.C. processed and analysed the Fermi/GBM data. R.-C.C., J.Y., C.-W.W., W.-J.T. and S.-L.X. processed and analysed the GECAM data, investigated the statistical methods, performed the periodicity search, and conducted the validation and significance estimation of the QPO. R.-C.C. and J.Y. investigated and applied the Bayesian inference method and performed the power spectral fitting for the QPO. R.-C.C. conducted the temporal and spectral analysis of the QPO. B.Z. and B.-B.Z. provided the idea for the physical interpretation of the QPO. R.-C.C., Y.-H.I.Y., J.Y., B.-B.Z. and B.Z. investigated the physical model, produced the schematic diagram and contributed to the discussion of the physical implications. R.-C.C., B.-B.Z., B.Z., C.-W.W., W.-J.T, S.-L.X., J.Y., Y.-H.I.Y. and S.-N.Z. contributed to discussions about the results. R.-C.C., B.-B.Z. and B.Z. wrote the paper, with contributions from all authors.

\item[Competing Interests] The authors declare no competing interests.

\item[Additional information] Correspondence and requests for materials should be addressed to Bin-Bin Zhang, Shao-Lin Xiong or Bing Zhang.

\end{addendum}

\clearpage
\setcounter{figure}{0}
\setcounter{table}{0}
\captionsetup[table]{name={\bf Extended Data Table}}
\captionsetup[figure]{name={\bf Extended Data Fig.}}

\clearpage
\begin{figure}
\centering
\includegraphics[width=\textwidth]{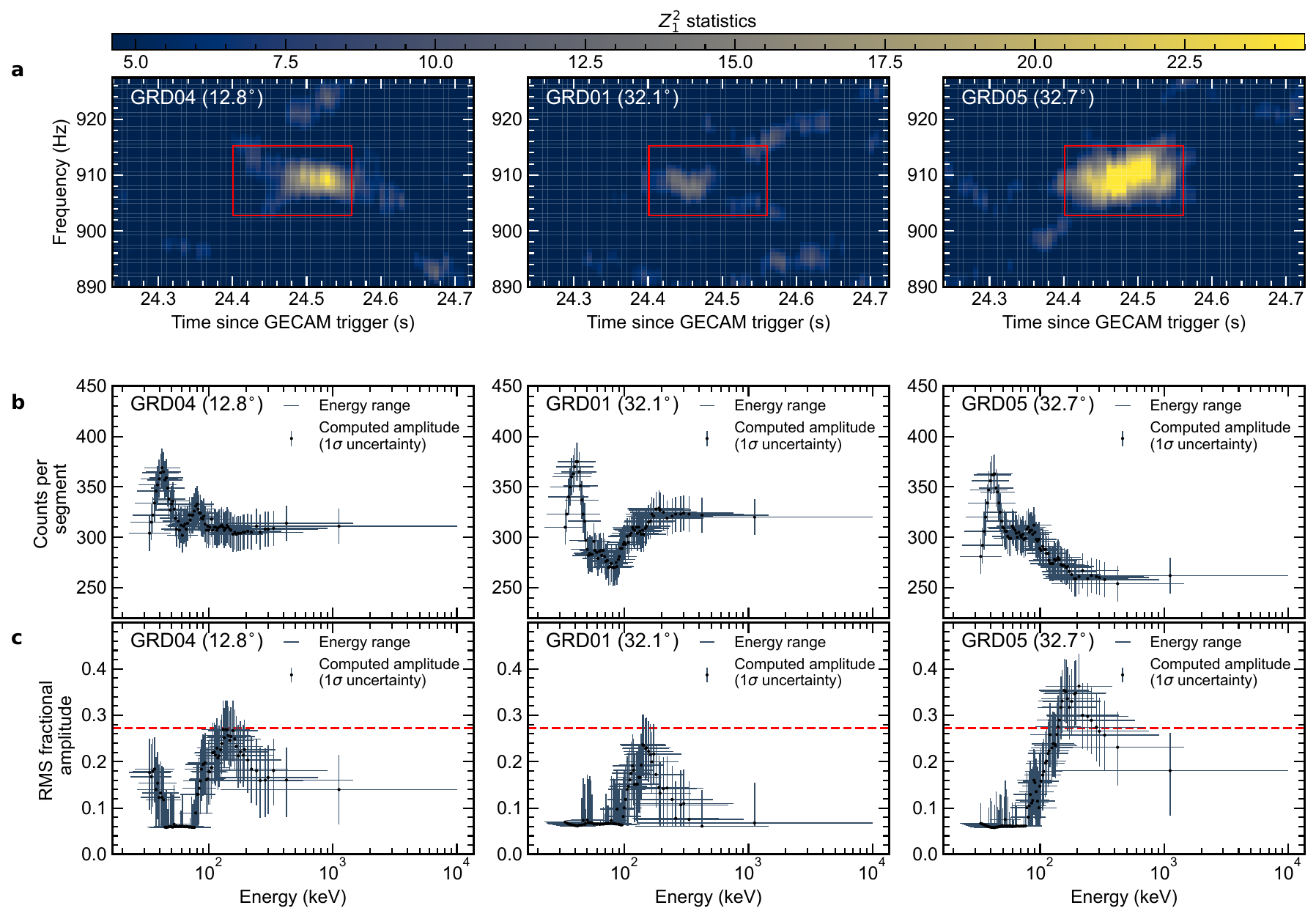}
\caption{\noindent\textbf{Consistent detection of the 909-Hz QPO across multiple detectors on GECAM-B.}
The incident angles to GRB~230307A for each detector are presented in brackets following the detector names.
\textbf{a,} Dynamical power spectra based on \( Z^2_1 \) statistics calculated from overlapping 160-ms intervals spaced at 7~ms. TTE data are extracted within the energy range of 98--248~keV from each detector individually. The red box marks the time interval of [24.401, 24.561]~s from \( T_0 \) and \( 909 \pm 1/0.16 \)~Hz. Each dynamical power spectra has the same lower threshold \( Z^2_1 \) statistic of 90\% confidence level in \( \chi^2_2 \) distribution.
\textbf{b,} Photon counts within [24.401, 24.561]~s in different energy ranges. See Extended Data Table \ref{etab1} for details.
\textbf{c,} r.m.s. fractional amplitudes at 909 Hz. Measurements from the \( Z^2_3 \) statistics are marked with blue error bars within the same time and energy interval as in \textbf{b}.
The QPO is consistently observed across different detectors, each showing similar energy dependence and a peak r.m.s. fractional amplitude of 27.2\% (marked by red dashed lines), indicating that the QPO is captured independently by each detector.}
\label{efig1}
\end{figure}

\clearpage
\begin{figure}
\centering
\includegraphics[width=0.7\textwidth]{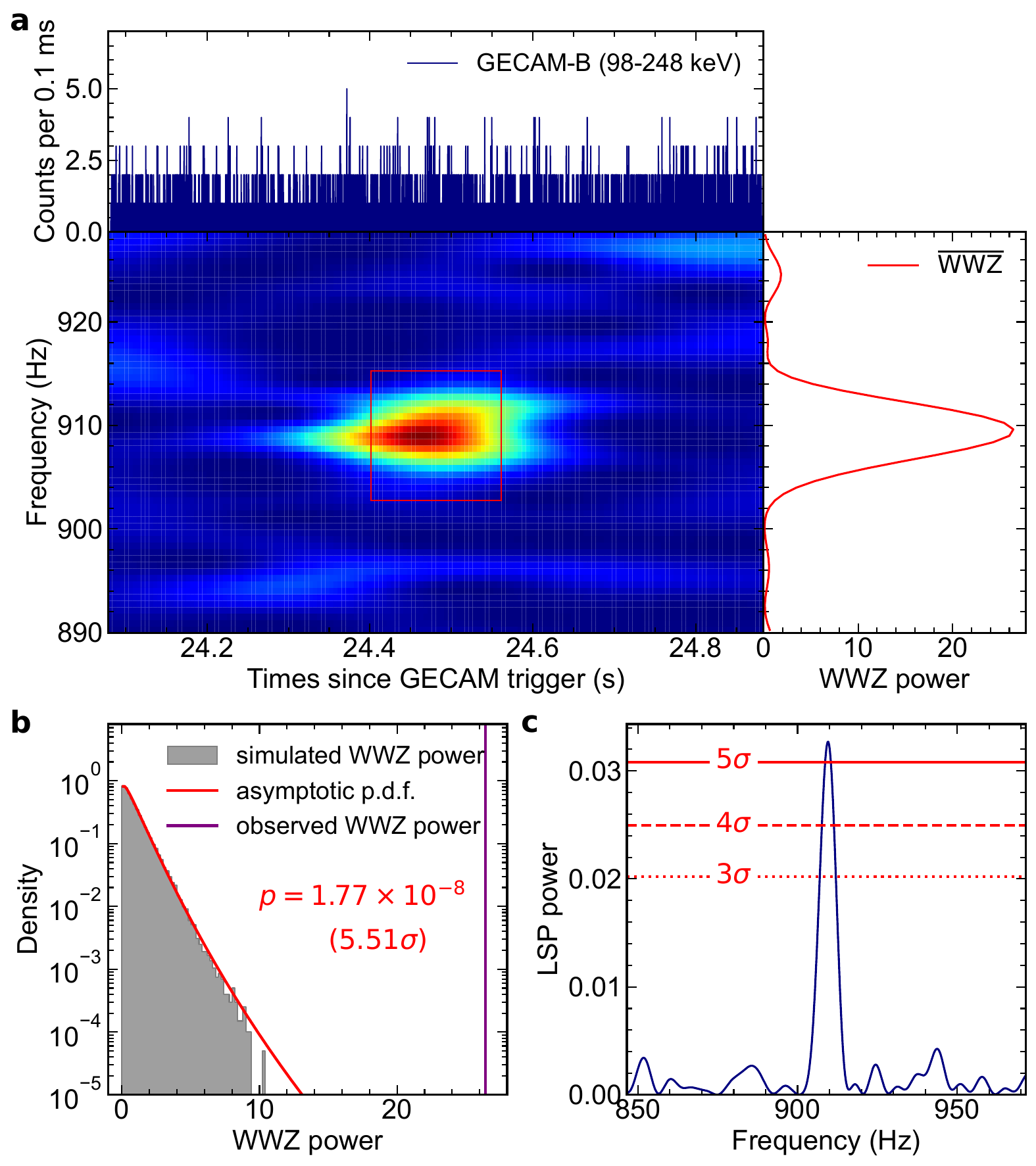}
\caption{\noindent\textbf{Monte Carlo simulations to verify the 909-Hz signal in GRB~230307A}
\textbf{a,} The upper panel shows the GECAM-B light curve of GRB~230307A from [24.081, 24.881]~s after $T_0$, binned at 0.1~ms, with no notable pulses or data corruption detected around the QPO interval ([24.401, 24.561]~s from $T_0$). The median panel displays the WWZ spectrogram, where the red box marks the QPO interval and the associated frequency range ($909 \pm 1/0.16$~Hz). The right panel shows the averaged WWZ power spectrum for the QPO interval.
\textbf{b,} Distribution of simulated WWZ power from synthetic light curves generated based on the observed light curve within the QPO interval. The red curve approximates the probability density function using an F-distribution, while the purple vertical line indicates the observed power of the 909-Hz signal.
\textbf{c,} Lomb-Scargle periodogram (LSP) of the light curve within the QPO interval. The red dotted, dashed, and solid lines represent the $3\sigma$, $4\sigma$, and $5\sigma$ false alarm levels, calculated using the Baluev method\cite{2008MNRAS.385.1279B}.}
\label{efig2}
\end{figure}

\clearpage
\begin{figure}
\centering
\includegraphics[width=0.9\textwidth]{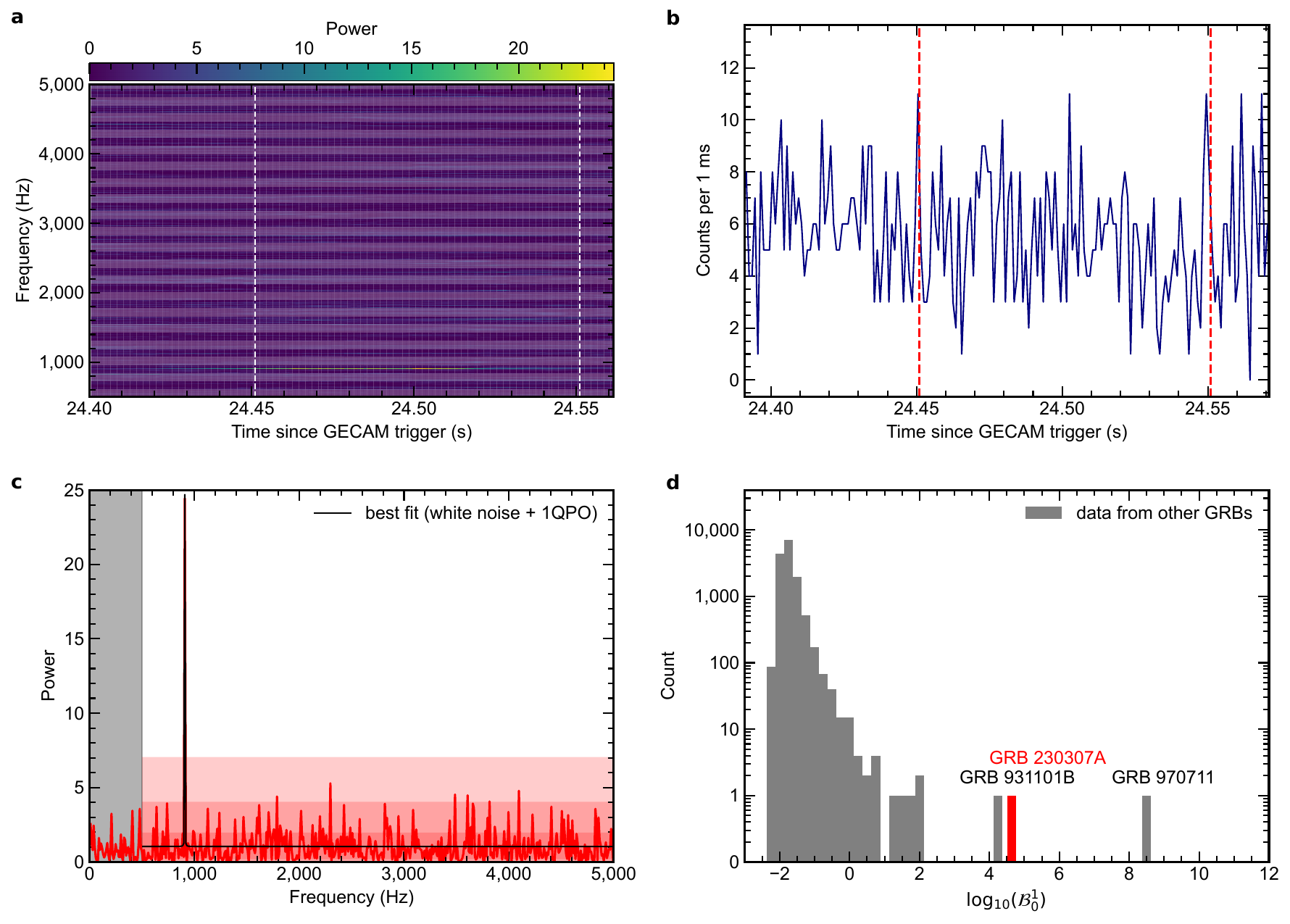}
\caption{\noindent\textbf{Bayesian model fitting for the 909-Hz QPO in GRB~230307A.} 
\textbf{a,} Dynamical power spectra based on the Groth normalized power spectra\cite{1975ApJS...29..285G}, mapped using overlapping 100-ms intervals with a step size of 1 ms. The white vertical line marks the central time of the interval where the maximal power at 909 Hz was detected.
\textbf{b,} Light curve of GRB~230307A. The red dashed vertical lines indicate the selected time interval [24.451, 24.551] s from $T_0$ used to generate the PDS.
\textbf{c,} PDS with the best-fit QPO model. The gray shaded area indicates the red noise region within the 10--500~Hz range, showing no notable red noise excess. The red bands denote the $1\sigma$, $2\sigma$, and $3\sigma$ confidence intervals under the assumption of white noise only. The black curve represents the expected power spectrum for the model combining white noise and one QPO.
\textbf{d,} Differential distribution of the Bayes factors, illustrating the ratio of Bayesian evidence for PDS models with and without a QPO. The Bayes factor $\mathcal{B}^1_0$ represents the ratio between the Bayesian evidence for the white noise plus one QPO model and the white noise model. All subsets were calculated using a 0.1-s segment length and a frequency range of 500--5,000~Hz. The gray histogram shows data from other short GRBs\cite{2023Natur.613..253C}, while the red histogram highlights the Bayes factors of the 909~Hz signal within [24.451, 24.551]~s from $T_0$ in GRB~230307A, comparable to those of GRB~931101B and GRB~970711\cite{2023Natur.613..253C}.}
\label{efig3}
\end{figure}

\clearpage
\begin{figure}
\centering
\includegraphics[width=0.6\textwidth]{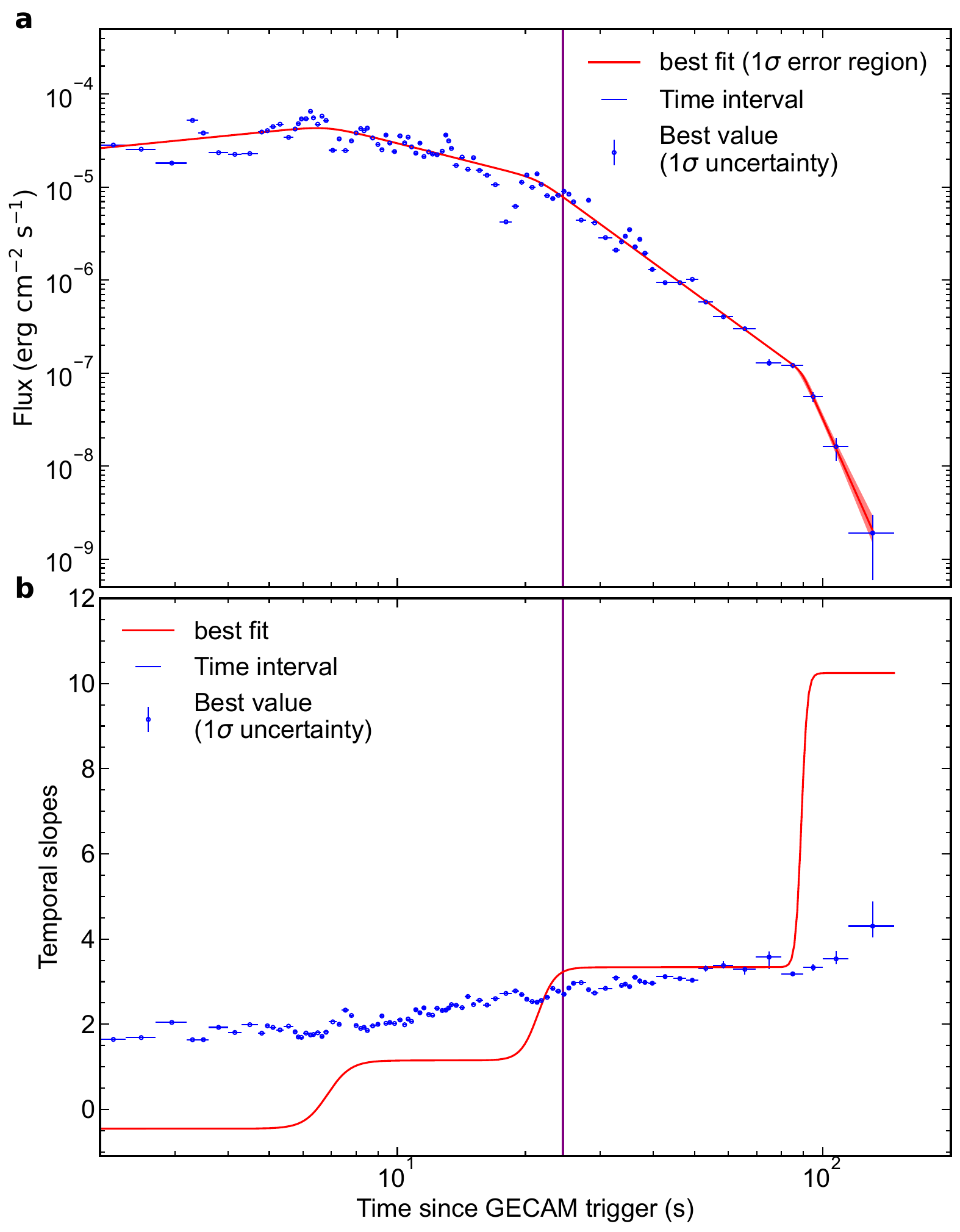}
\caption{\noindent\textbf{Coincidence between the 909-Hz signal detection interval and the transition to high-latitude emission.}
\textbf{a,} Flux evolution of GRB~230307A based on GECAM data in 98--248~keV band. Blue error bars represent the time-resolved flux estimated from the best-fit spectral fitting results detailed in ref.\cite{2023arXiv230705689S}, including the corresponding time intervals and 1$\sigma$ uncertainties. The red curve shows the best-fit smoothly broken power law with the corresponding 1$\sigma$ error region.
\textbf{b,} Evolution of temporal slopes. Blue points indicate the temporal slopes predicted by the ``curvature effect'', and the red curve represents the slopes derived from the smoothly broken power law fitting in \textbf{a}.
The vertical purple band marks the interval of 909-Hz signal detection, which coincides with the transition point where the ``curvature effect'' begins to dominate the observed gamma-ray flux.}
\label{efig4}
\end{figure}

\clearpage
\begin{figure}
\centering
\includegraphics[width=0.7\textwidth]{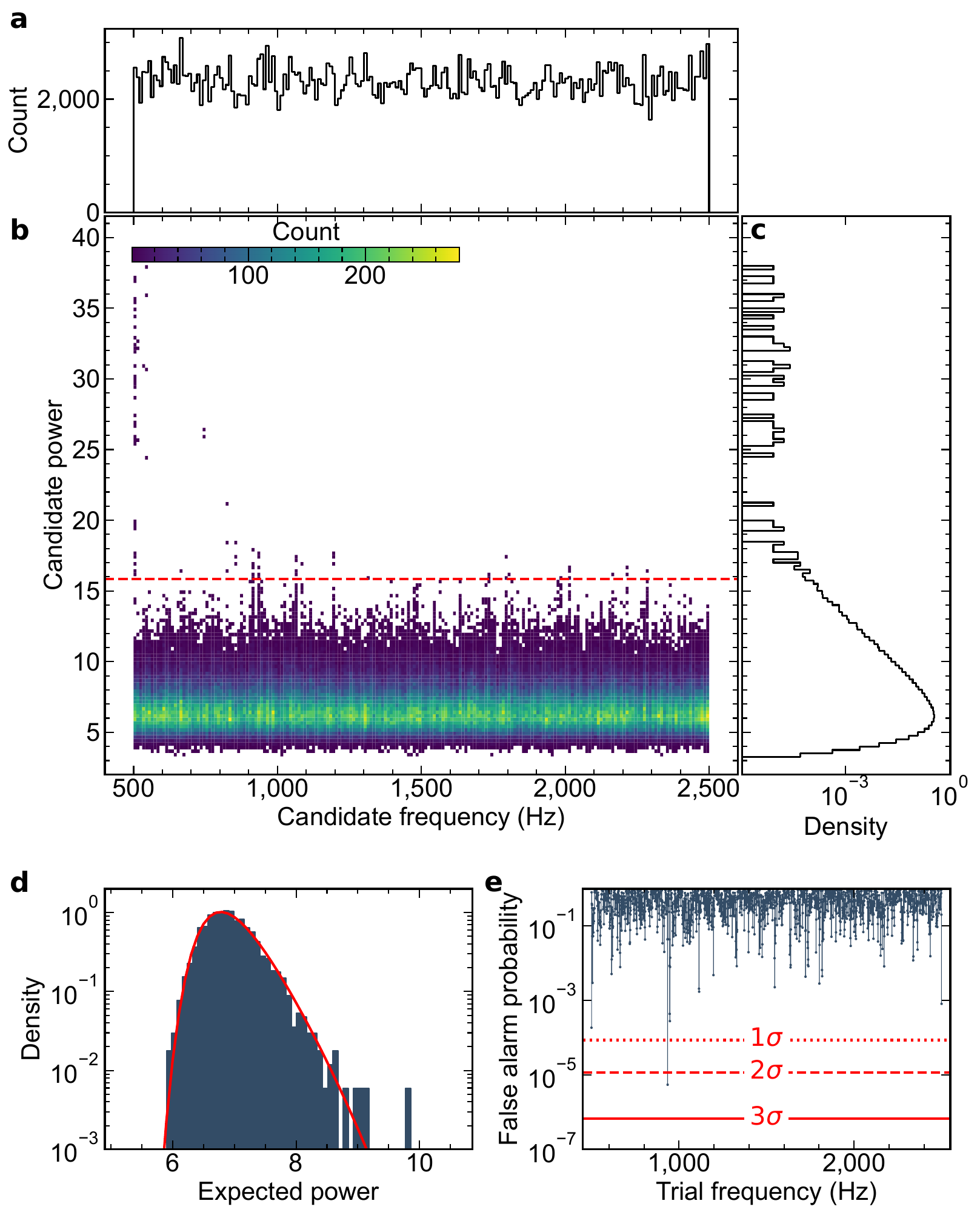}
\caption{\noindent\textbf{Result of the blind search for periodicity in GRB~211211A.}
We obtained candidates from 462,852 subsets of event data, partitioned from 8,935 segments, covering the light curve from [-0.016, 81.910]~s from $T_{0,{\rm 211211A}}$ in the 5--975~keV range from detectors n2 and na onboard Fermi/GBM.
\textbf{a, b, c, d, e,} Similar to Fig. \ref{fig1}, but for GRB~211211A. The maximum candidate power was observed at 546~Hz in \textbf{b}, corresponding to a signal within the time interval [14.880, 15.075]~s from $T_{0,{\rm 211211A}}$. This signal exhibited powers exceeding the threshold of $R_{\rm false} = 15.85$ in the 500--549~Hz range, with a representative FAP of approximately $31.09\%$, indicating that the signal is likely due to noise. Another signal within [50.914, 51.029]~s from $T_{0,{\rm 211211A}}$ showed power exceeding the threshold at 935 and 936~Hz, but the representative FAP of about $1.09\%$ renders this signal insignificant.}
\label{efig5}
\end{figure}

\clearpage
\begin{figure}
\centering
\includegraphics[width=0.9\textwidth]{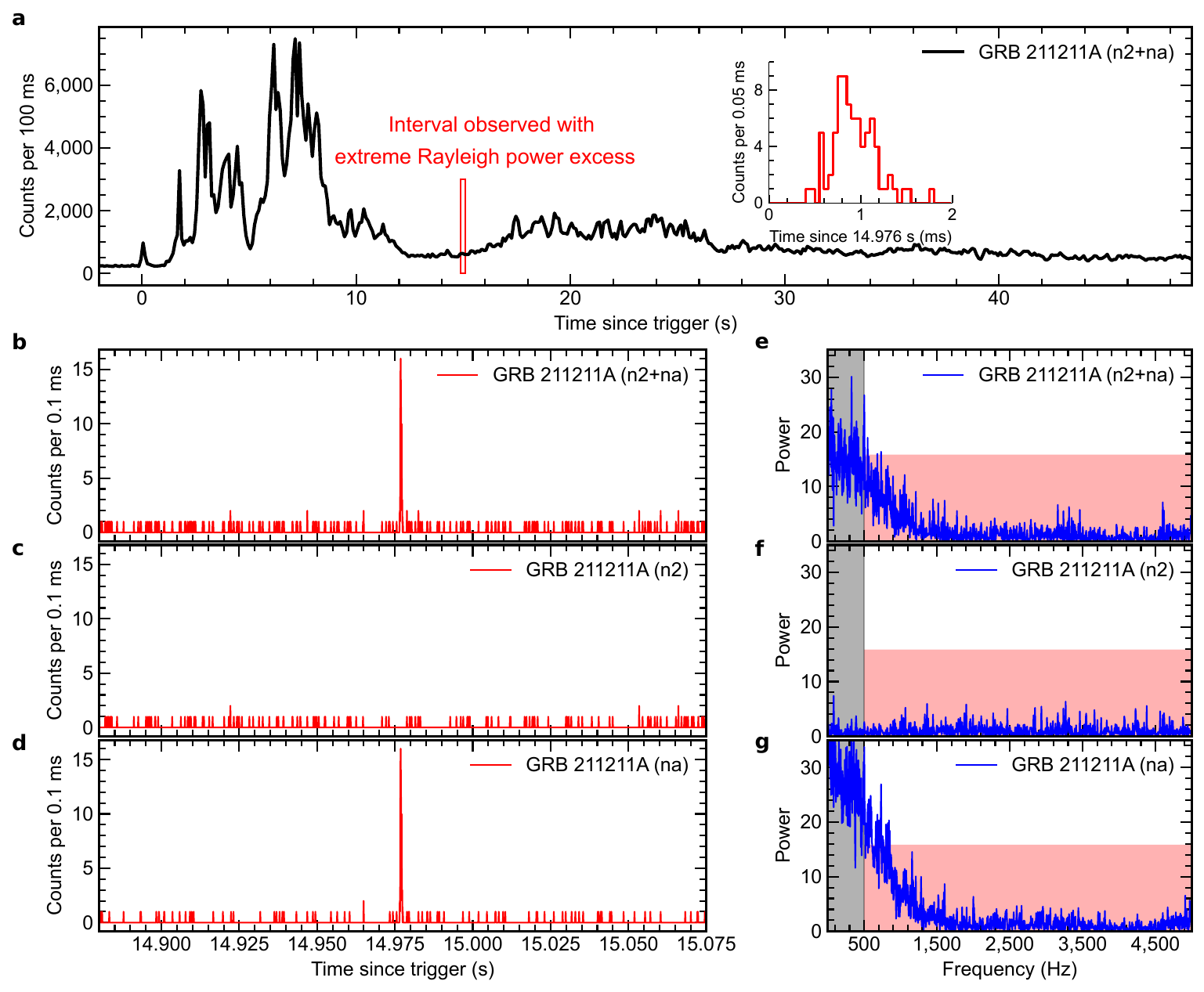}
\caption{\textbf{Very strong red noise during GRB~211211A.}
\textbf{a,} Light curve of GRB~211211A from Fermi/GBM in 5--975~keV. As shown in Extended Data Fig.~\ref{efig5}, a signal yielding candidate powers exceeding the threshold within 500--549~Hz was observed. The corresponding time interval is marked with a red box. The inset shows a detailed view of this signal, revealing a very bright spike lasting about 1~ms.
\textbf{b, c, d,} Light curves in the 5--20~keV range from Fermi/GBM detectors with incident angles less than 60$^\circ$. The spike, observed at around 14.977~s, was detected only by the NaI detector na (50.03$^\circ$) and not by detector n2 (28.37$^\circ$). Given the spike's absence in detector n2 and its predominant contribution in the 5--20~keV range, it was excluded as a hyper flare from GRB~211211A.
\textbf{e, f, g,} Groth-normalized\cite{1975ApJS...29..285G} power spectra corresponding to the light curves in \textbf{b, c, d}. The grey shaded area denotes the 0--500~Hz frequency range, while the red shaded area highlights the 500--5,000~Hz region and the threshold power from Extended Data Fig. \ref{efig5} \textbf{a}. The bright spike detected by the na detector produced a series of red noise excesses extending up to 2,000~Hz, leading to the observation of extreme Rayleigh power during the search.}
\label{efig6}
\end{figure}

\clearpage
\begin{table*}
\centering
\scriptsize
\caption{\textbf{Energy Segmentation for the Time Interval [24.401, 24.561]~s from $T_0$ Using GECAM-B Data.} Each energy range corresponds to an energy channel range with a photon count comparable to that of channels [99, 168], which span approximately 98--248~keV. The boundaries of each energy range are defined by the physical energy range of the corresponding energy channel ranges.}
\label{etab1}
\begin{tabular}{llllll}
\toprule
\multicolumn{6}{c}{\textbf{Energy Ranges (keV)}} \\
\midrule
22--48 & 23--50 & 24--51 & 25--52 & 26--53 & 27--54 \\
28--56 & 29--57 & 29--58 & 30--59 & 31--60 & 32--62 \\
33--63 & 34--64 & 35--66 & 37--67 & 38--68 & 39--70 \\
40--71 & 41--72 & 42--75 & 43--77 & 44--79 & 45--81 \\
46--84 & 47--85 & 48--87 & 50--90 & 51--93 & 52--95 \\
53--98 & 54--100 & 56--103 & 57--106 & 58--110 & 59--112 \\
60--113 & 62--117 & 63--119 & 64--122 & 66--124 & 67--128 \\
68--132 & 70--136 & 71--138 & 72--142 & 74--148 & 75--152 \\
77--158 & 78--162 & 79--168 & 81--170 & 82--175 & 84--183 \\
85--187 & 87--194 & 88--198 & 90--205 & 92--212 & 93--217 \\
95--228 & 96--233 & 98--248 & 100--261 & 101--271 & 103--286 \\
105--305 & 106--335 & 108--345 & 110--383 & 112--432 & 113--514 \\
115--579 & 117--688 & 119--755 & 121--910 & 122--1448 & 124--10053 \\
\bottomrule
\end{tabular}
\end{table*}

\clearpage
\begin{table*}
\centering
\begin{minipage}{\textwidth}
\caption{\textbf{Results of PDS Model Fitting.} The best-fit parameters are derived from the optimal values corresponding to the maximum likelihood estimates of the posterior distributions, as determined by the \texttt{MULTINEST} algorithm. The upper and lower bounds indicate the 1$\sigma$ uncertainties.}
\label{etab2}
\begin{tabular}{lcccccc}
\toprule
\textbf{Model} & \boldmath{$A_{\rm white}$} & \boldmath{$A_{\rm QPO}$} & \boldmath{$f_0$} & \boldmath{$\Delta f_0$} & \boldmath{$\ln\mathcal{L}$} & \boldmath{$\ln E$} \\
\midrule
White noise only & $0.07^{+0.05}_{-0.04}$ & - & - & - & $-479.07$ & $-483.04$ \\
White noise plus QPO & $0.04^{+0.04}_{-0.03}$ & $23.65^{+4.23}_{-5.43}$ & $909.08^{+0.48}_{-0.51}$ & $0.95^{+1.95}_{-0.40}$ & $-458.48$ & $-472.37$ \\
\bottomrule
\end{tabular}
\end{minipage}
\end{table*}

\end{document}